\documentclass[%
 aip,
 amsmath,amssymb,
 reprint,%
]{revtex4-1}

\usepackage{graphicx}
\usepackage{dcolumn}
\usepackage{bm}

\usepackage[utf8]{inputenc}
\usepackage[T1]{fontenc}
\usepackage{mathptmx}
\usepackage{xcolor}
\usepackage{float}

\begin{document}

\preprint{AIP/123-QED}

\title[]{Mechanical investigations of free-standing SiN membranes patterned with one-dimensional photonic crystal structures}

\author{Ali Akbar Darki}
\affiliation{Department of Physics and Astronomy, Aarhus University, DK-8000 Aarhus C, Denmark}
\author{Robin Vinther Nielsen}
\affiliation{Department of Physics and Astronomy, Aarhus University, DK-8000 Aarhus C, Denmark}
\author{Jens Vinge Nygaard}
\affiliation{Department of Biotechnological and Chemical Engineering, Aarhus University, DK-8000 Aarhus C, Denmark}
\author{Aur\'{e}lien Dantan}
\email{dantan@phys.au.dk}
\affiliation{Department of Physics and Astronomy, Aarhus University, DK-8000 Aarhus C, Denmark}

\date{\today}

\begin{abstract}
A detailed investigation of the structural and vibrational properties of various prestressed silicon nitride membranes patterned with one-dimensional photonic crystal structures is presented. The tensile stress-related deformation of the structure in the vicinity of the patterned area is determined by Atomic Force Microscopy scans, while the resonance frequencies and quality factors of the out-of-plane membrane vibrations are measured using optical interferometry.
We show that these noninvasive measurements, combined with results of finite element simulations, provide accurate knowledge on the tensile stress, the elasticity modulus and the density of these nanostructured thin films. The obtained results are interesting in two ways: first, they show that such highly reflective thin membranes already exploited in various photonics applications possess high-mechanical quality, which also makes them attractive for optomechanics and sensing applications. Second, they represent a nondestructive method to determine key material parameters which can be applicable to a broad range of fragile nanostructured thin films.
\end{abstract}

\maketitle

\section{Introduction}

Due to its structural, optical, electrical and chemical properties, silicon nitride (SiN) is a ubiquitious material for countless applications within photonics, Micro Electro-Mechanical Systems (MEMS) technology, sensing, etc. Progress in fabrication technology has enabled the realization of numerous miniaturized SiN devices at the micro- and nanoscales. The material properties of thin films and nanostructures, however, are often quite different from those of their bulk counterparts and they can substantially vary depending on the fabrication/deposition methods used~\cite{Kaloyeros2016,Kaloyeros2020,Hegedus2021}. For suspended films, the knowledge of mechanical quantities, such as the film density, its elasticity modulus, its residual stress level, etc., is essential to proper design, fabrication and application of devices in which, e.g., the vibrations of the structures are exploited. Futhermore, nanostructuring of films performed to tailor the optical or electrical response typically modifies the material on lengths scales which can dramatically affect its mechanical properties. 

An emblematic example, which we will focus onto in this work, consists of a suspended SiN thin film, patterned with a subwavelength structure in order to tailor its optical or mechanical response, and widely applied in applications for e.g. lasing~\cite{ChangHasnain2012}, flat optics~\cite{Capasso2014}, sensing~\cite{Wang1993,Quaranta2018}, optomechanics~\cite{Kemiktarak2012,Bui2012,Kemiktarak2012a,Yu2014,Norte2016,Reinhardt2016,Chen2017,Tsaturyan2017,Moura2018,Cernotik2019}. A number of methods for characterizing the material properties of thin films~\cite{Kraft2001,Abadias2018}--nanoindentation, tensile and bulge testing, point deflection, etc.--have been applied to SiN films~\cite{Kaloyeros2016,Kaloyeros2020,Hegedus2021,Tabata1989,Taylor1991,Stewart1991,Gardeniers1996,Yen2003,Chuang2004,Kaushik2005,Huang2005,Kim2006,Huang2006,Martins2009,Hwangbo2012,Huszank2016,Capelle2017,Jugade2021}. However, many of these methods are destructive and not necessarily well-suited for the study of devices where the preservation of high-quality mechanical resonances is essential. As abovementioned, key parameters, such as the SiN film density, its elasticity modulus and level of residual stress, typically vary substantially depending on the many parameters of the fabrication recipe~\cite{Kaloyeros2016,Kaloyeros2020}. It is thus desirable to use noninvasive methods to gain accurate insight into post-fabrication material properties of fragile thin films. Such an insight can be used in further modelling and accounting of their performances or for further optimization of their design and fabrication~\cite{Hoej2021,Serra2021,Conte2022}.

In this work, we report on investigations of mechanical properties of prestressed silicon nitride membranes patterned with one-dimensional photonic crystal structures~\cite{Nair2019}. Atomic Force Microscopy (AFM) scans are applied to evaluate the tensile stress-related deformation of the structure in the vicinity of the patterned area~\cite{Darki2021}, which essentially depend on the ratio of the elasticity modulus and the pretension level of the membrane. Optical interferometric measurements of the out-of-plane membrane vibrations enables determination of their mechanical resonance frequencies and quality factors, which are essentially dictated by the ratio of the tensile stress and the density of the film. Finally, AFM point deflection spectroscopy is applied to extract the tensile-stress dominated spring constant of the membranes. Combined with the results of finite element simulations, we show that these three measurements allow for noninvasively determining the state of residual stress, the elasticity modulus and the density of the nanostructured films. The method is applied to membranes possessing different levels of prestress obtained when varying film thickness and photonic crystal structure patterning.

The obtained results are interesting in two ways: first, they specifically show that these highly reflective thin membranes already exploited in various photonics applications~\cite{Parthenopoulos2021,Toftvandborg2021,Darki2021b} also possess high-mechanical quality, which makes them attractive for sensing~\cite{Moura2018,Naesby2017,Naesby2018} and for cavity optomechanics with single~\cite{Thompson2008,Wilson2009,Kemiktarak2012a,Chen2017,Hoej2021} or multiple~\cite{Xuereb2012,Nair2017,Gartner2018,Piergentili2018,Wei2019,Yang2020b,Fitzgerald2021} SiN membrane resonators. Secondly, the techniques used, which exploit the nanostructuring-related stress-induced structural deformations, provide a general method for nondestructively and accurately determining key material parameters of a broad range of fragile nanostructured thin films.

The paper is organized as follows: three types of samples and the investigative methods are presented in Sec.~\ref{sec:samples_methods}, measurement results and their analysis for the three batches are detailed in Sec.~\ref{sec:results} and a conclusion is given in Sec.~\ref{sec:conclusion}.

 
\section{Samples and characterization methods}\label{sec:samples_methods}

\subsection{Samples}\label{sec:samples}

Three groups of samples are investigated in this work:

\begin{itemize} 

\item \underline{Batch A}: low-stress ($\sim 0.1$ GPa), $315$ nm-thick, 500 $\mu$m-square SiN membranes custom-produced by Norcada, Inc.~\cite{Norcada} with a 200 $\mu$m-square grating patterned at the center of the suspended film. The gratings have essentially the same period ($685$ nm) and fill ratio ($0.57$), and slightly different finger depths (225-235 nm).  The films are suspended on a 500 $\mu$m-thick, 5 mm-square Si frame.

\item \underline{Batch B}: low-stress ($\sim 0.1$ GPa), $200$ nm-thick, 500 $\mu$m-square SiN membranes patterned with a 200 $\mu$m-square grating. The bare films were purchased from Norcada and the gratings were subsequently patterned using Electron Beam Lithography (EBL) and chemical etching following the recipe described in~\cite{Nair2019,Parthenopoulos2021}.  The gratings have similar periods ($\sim 855$ nm) and fill ratios ($\sim 0.5$), but substantially different finger depths (50-180 nm). The films are suspended on a 200 $\mu$m-thick, 5 mm-square Si frame.

\item \underline{Batch C}: high-stress ($\sim 1$ GPa), $200$ nm-thick, 500 $\mu$m-square Si$_3$N$_4$ membranes patterned following the same recipe as batch B, but with square gratings of different sizes ($100-400$ $\mu$m) and different geometrical parameters (see Tab.~\ref{tab:C} in Sec.~\ref{sec:C}). The films are suspended on a 200 $\mu$m-thick, 5 mm-square Si frame.

\end{itemize}

\begin{figure}
\includegraphics[width=\columnwidth]{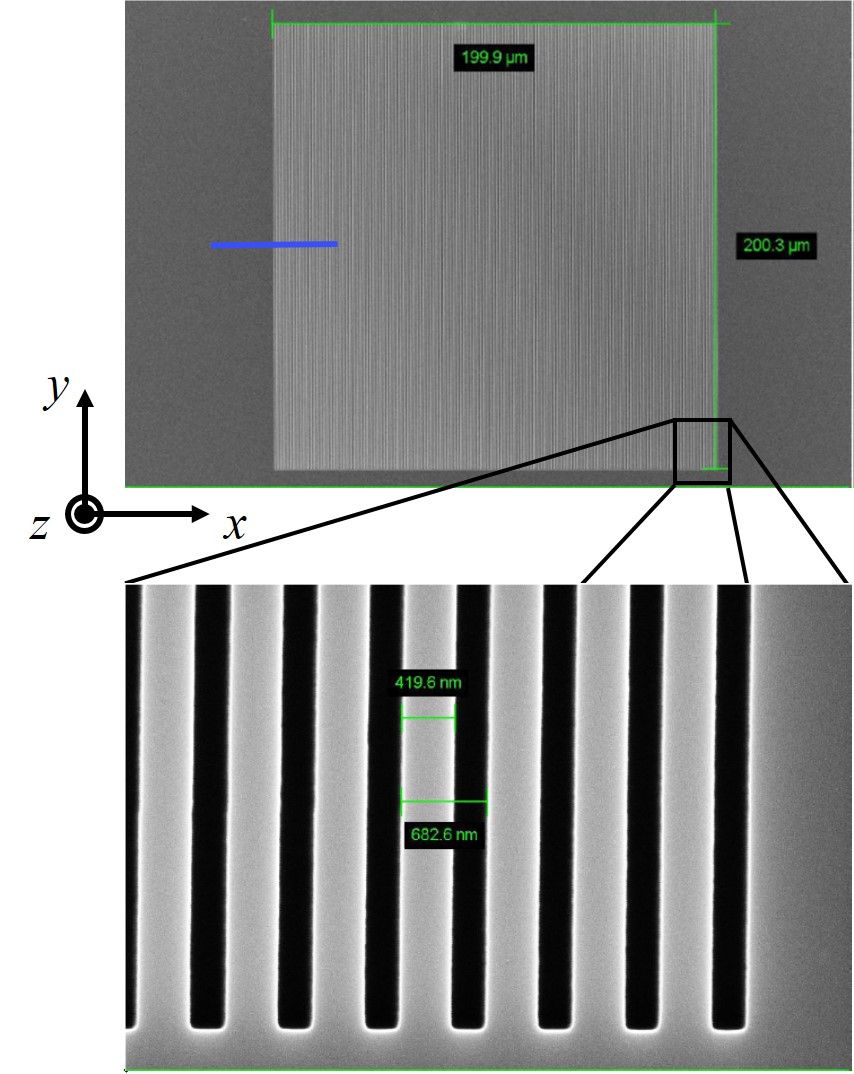}
\caption{SEM topview pictures of a patterned membrane from batch A (Courtesy of Norcada~\cite{Norcada}). The bottom picture is reproduced with permission from Opt. Express {\bf 30}, 3962 (2022). Copyright 2022 Optica Publishing Group. The horizontal blue line in the top picture indicates the AFM scan line used in Fig.~\ref{fig:afm}.}
\label{fig:norcada}
\end{figure}

Scanning Electron Microscope (SEM) pictures of a patterned membrane from batch A are shown in Fig.~\ref{fig:norcada}. The optical properties of these nanostructured membranes, and specific applications within photonics, have been investigated in detail in~\cite{Nair2019,Parthenopoulos2021,Darki2021,Toftvandborg2021,Darki2021b}. The structural deformations of some samples from batch $C$ have been investigated in~\cite{Darki2021} using AFM profilometry. Let us point out that, while we distinguish the batches by their "low" and "high" stress levels, tension effects strongly dominate over the flexural rigidity of the films in all cases. In the next section we apply the AFM profilometry method to the precise characterization of the deformations of all these highly-prestressed patterned structures.

\subsection{Topological characterization}

\begin{figure}[h]
\includegraphics[width=\columnwidth]{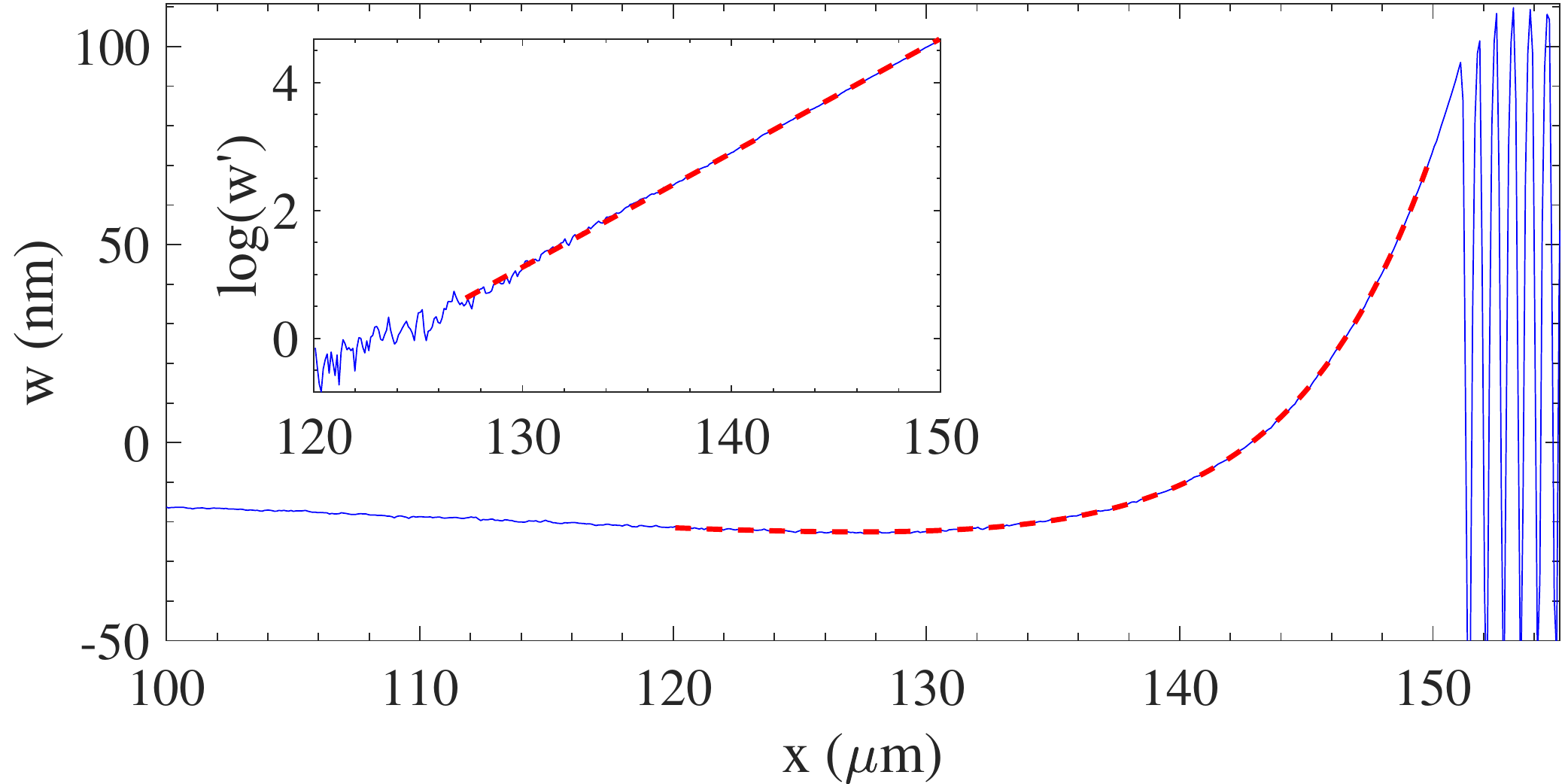}
\caption{Result of an AFM scan in the direction perpendicular to the grating fingers and in the middle of the patterned area border for sample A2 (horizontal blue line in Fig.~\ref{fig:norcada}). The red dashed line shows the result of a fit with the expected theoretical profile. The inset shows on a logarithmic scale the deformation $w'$, corrected from the pre-edge depression (second-order polynomial~\cite{Darki2021}). The red line shows the result of a fit to the upward deflection part.
}
\label{fig:afm}
\end{figure}

In order to characterize the topology of the nanostructured membranes we employ the AFM scanning method introduced in~\cite{Darki2021}. In brief, a Brucker Dimension Edge AFM using RTESPA-300 tips is used to scan the vertical deflection, $w$, in the middle of the border of the patterned area in the $x$-direction, perpendicular to the grating fingers. The result of such a deflection measurement for sample A2 is shown in Fig.~\ref{fig:afm}; it clearly demonstrates the deflection of the thin film due to the abrupt change in its thickness at the border and the change in the stress distribution of the suspended film.

\subsubsection{Principle: infinite (1D) membrane uniformly etched in its center}

\begin{figure}
\includegraphics[width=\columnwidth]{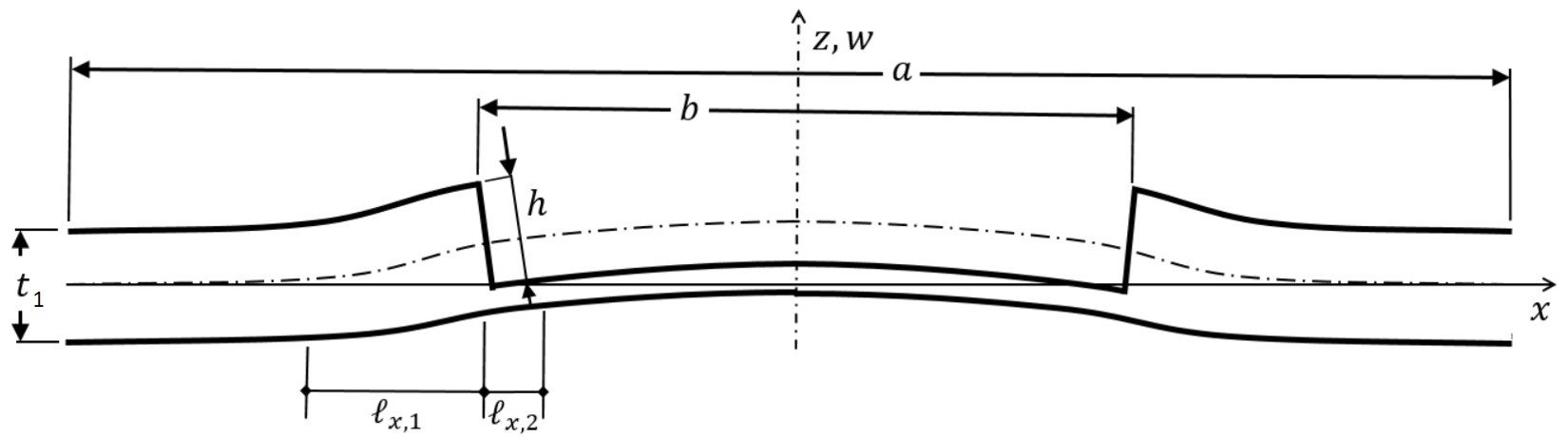}
\caption{Cross-section schematic of the deformation of a 1D membrane with width $a$ and whose thickness $t_1$ is reduced by $h$ in its middle section over a width $b$. $\ell_{x,1}$ and $\ell_{x,2}$ indicate the characteristic lengths of the vertical deflections outside and inside of the etched area, respectively.}
\label{fig:1D}
\end{figure}

This structural deformation was investigated in~\cite{Darki2021} and can be simply understood on the basis of a 1D model considering a thin plate with varying thickness, as depicted in Fig.~\ref{fig:1D}. Indeed, for a strongly pretensioned plate, which is infinitely long in the $y$-direction and whose thickness is reduced by $h$ in its central section in the $x$-direction, a positive deflection of the plate of about $h/2$ is expected to develop exponentially in the $z$-direction over a characteristic length determined by the ratio $E/\sigma_0$ of Young's elasticity modulus $E$ to the initial pre-stress level $\sigma_0$ of the suspended film~\cite{Darki2021}. 

More quantitatively, let us consider a plate of length $a$ in the $x$-direction, thickness $t_1$ and clamped at its edges, and assume that a height $h$ is removed from its central part over a length $b$, such that the thickness in this part is $t_2=t_1-h$. It was shown in~\cite{Darki2021} that the vertical deflections from the neutral line in the non-etched (1) and etched (2) parts of the membrane are of the form $w_i(x)=A_i+B_i\cosh(x/\ell_{x,i})$, ($i=1,2$), where $A_i$ and $B_i$ are constants and the characteristic deflection lengths $\ell_{x,i}$ are given by
\begin{equation}
\ell_{x,i}=\sqrt{\frac{Et_i^3}{12(1-\nu^2)N_0}},
\end{equation}
with $\nu$ being the Poisson ratio. For a pretensioned thin plate for which tension effects dominate over the flexural rigidity, the line force $N_0$ is, to a very good approximation, given by
\begin{equation}
N_0=\frac{\sigma_0}{\left(1-\frac{b}{a}\right)\frac{1}{t_1}+\frac{b}{a}\frac{1}{t_2}}.
\end{equation}
In such a 1D setting, measuring the deflection characteristic length(s) thus allows the determination of the ratio $E/\sigma_0$, provided the geometric parameters are known. Concommitantly, the stress is reduced in the non-etched (thick) part of the plate with respect to its initial value $\sigma_0$, while it is increased in the etched (thin) part. Close to the edges of the patterned area, the 1D model of Ref.~\cite{Darki2021} predicts simple relationships between these post-stress levels and the pre-stress level $\sigma_0$
\begin{equation}
\sigma_{x,1}\simeq\frac{\sigma_0}{1-\frac{b}{a}+\frac{b}{a}\frac{t_1}{t_2}} \hspace{0.2cm}\textrm{and}\hspace{0.2cm}
\sigma_{x,2}\simeq\frac{\sigma_0}{\left(1-\frac{b}{a}\right)\frac{t_2}{t_1}+\frac{b}{a}}.
\label{eq:sigmax}
\end{equation}

\subsubsection{General case: membrane patterned with 2D inhomogeneous structure}\label{sec:2Dmethod}

\begin{figure}[h]
\includegraphics[width=\columnwidth]{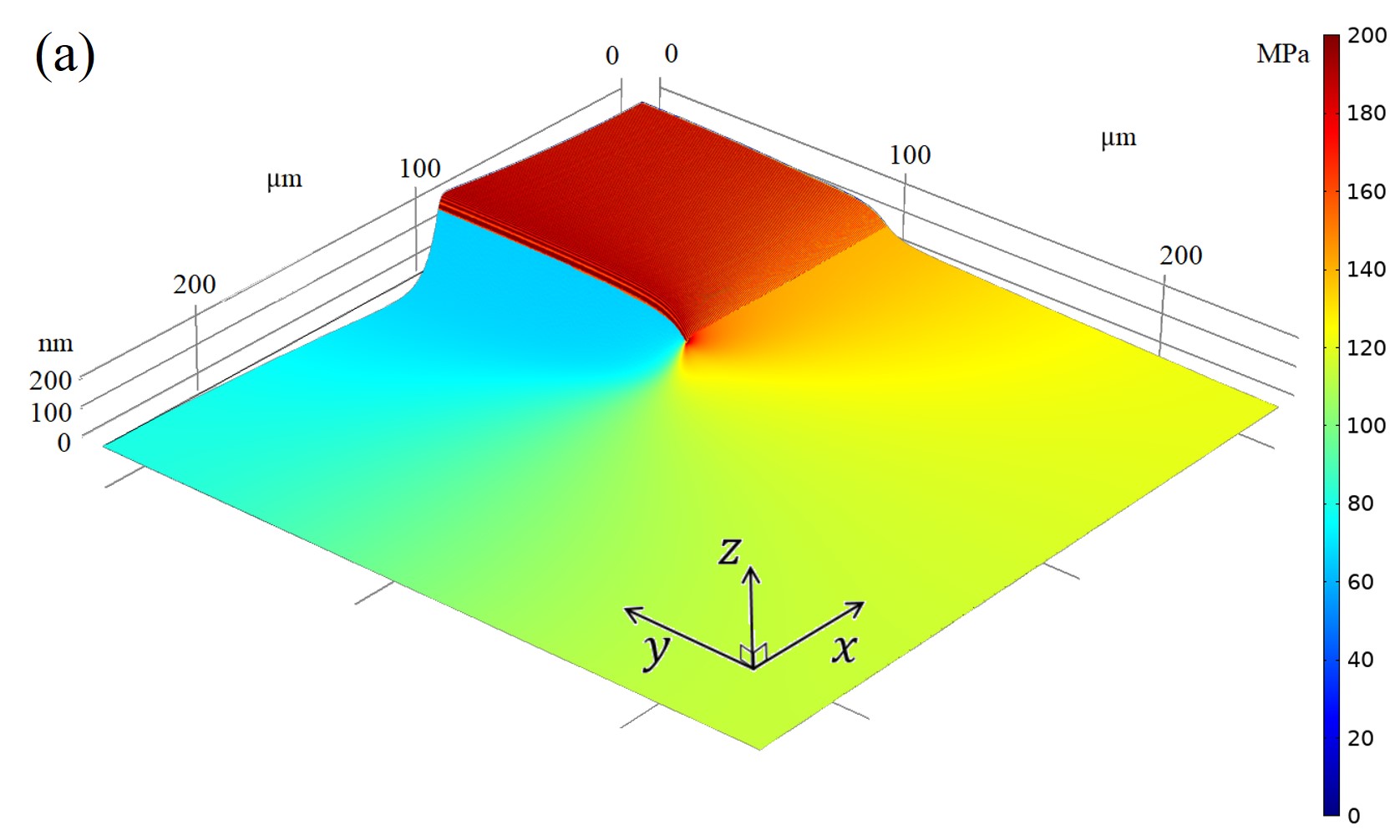}
\includegraphics[width=\columnwidth]{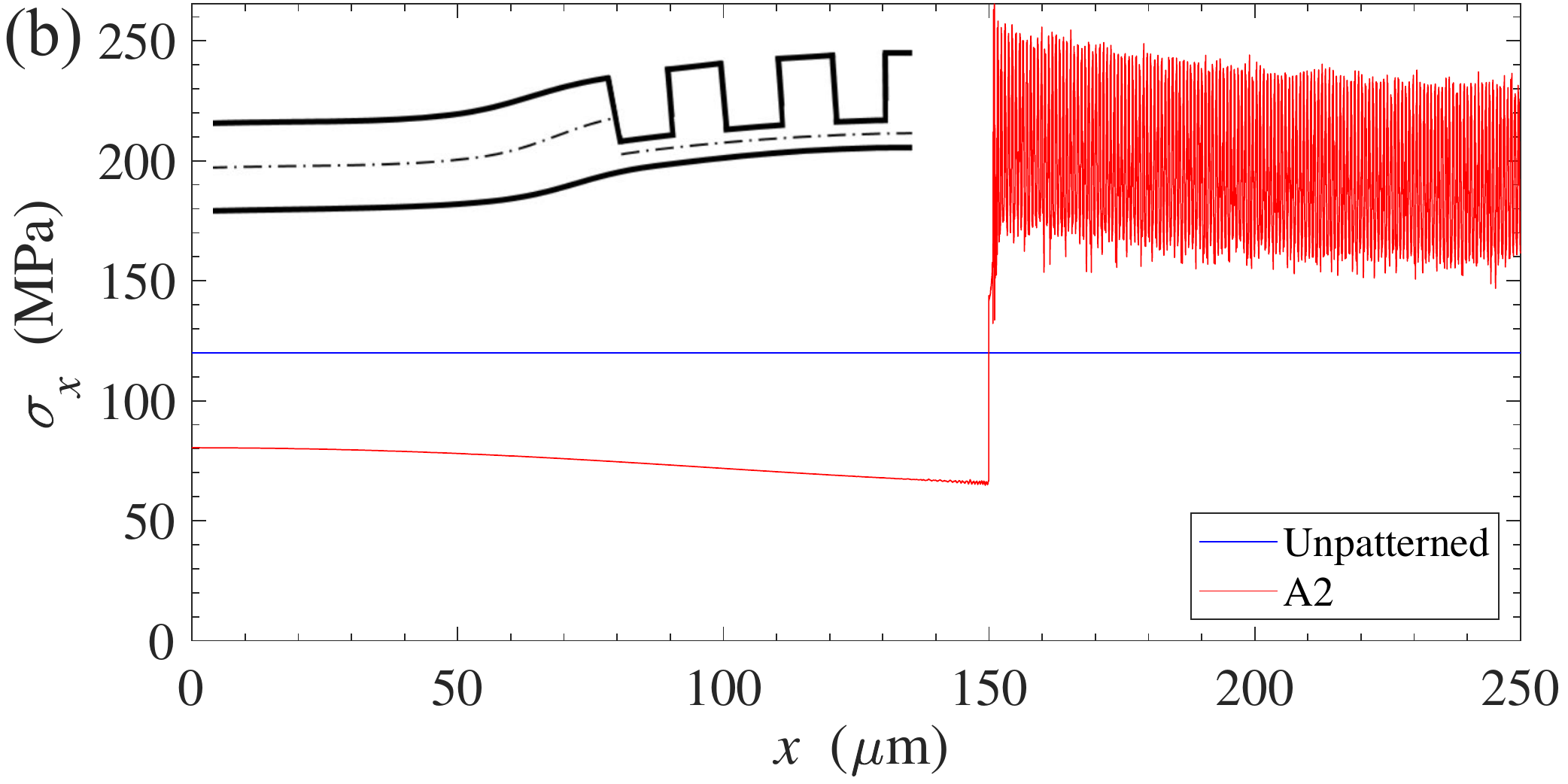}
\caption{(a) Results of FEM simulations of the vertical deflection in (a quarter of) sample A2 for a ratio $E/\sigma_0=2000$ and assuming $\sigma_0=120$ MPa. The colors give the corresponding simulated stress distribution. (b) Simulated variation along $x$ of the stress component $\sigma_x$ in the middle of the left finger. $\sigma_x$ is evaluated on the neutral line of the film in the unpatterned area and in the middle of the etched finger in the patterned area (dashed lines in the inset).}
\label{fig:comsol_stress}
\end{figure}

For a 2D plate patterned with a directionally asymmetric structure, such as a grating, the same behavior is qualitatively observed, but the stress distribution becomes inhomogeneous, as the stress components in the $x$- and $y$-directions are different and coupled together. While the 2D nature of the structure seemingly complicates the analysis, it is nonetheless possible to use the analytical findings of the 1D model, in particular, the deformation length $\ell_1$, to accurately evaluate the $E/\sigma_0$ ratio. Figure~\ref{fig:comsol_stress} shows the deflections and stress distributions resulting from simulations using the Finite Element Method (FEM)~\cite{Zienkiewicz2014} as implemented in COMSOL Multiphysics for sample A2 with $E/\sigma_0=2000$, $\sigma_0=120$ MPa and a Poisson ratio value of $\nu=0.27$. A stronger and sharper deflection is observed in the $x$-direction than in the $y$-direction, as the stress modification is stronger at edges parallel with the grating fingers than at the edges perpendicular to them. In order to match the maximal deflection in the $y$-direction at the center of the patterned area, the deflection in the $x$-direction has to slightly decrease before and after the abrupt increase at the edge. This is visible in both the measured (Fig.~\ref{fig:afm}) and simulated (Fig.~\ref{fig:comsol_stress}) profiles.

Scans such as the one shown in Fig.~\ref{fig:afm} are thus analyzed in the following way: first, the deflection profile is fitted with the sum of a second-order polynomial and an exponential function in the unpatterned region (Fig.~\ref{fig:afm}). The profile with the polynomial background subtracted is shown on a logarithmic scale in the inset of Fig.~\ref{fig:afm}, together with the result of a linear fit, which yields a value of $\ell_{x,1}=5.62\pm 0.04$ $\mu$m. Scans are typically performed on each edge parallel to the fingers for each sample. In principle, scans in the $y$-direction over the edges perpendicular to the grating fingers could also be used, but we limit ourselves here to scans in the $x$-direction.

In the previous 1D homogeneous geometry, the deflection lengths $\ell_{x,i}$ are simply related to the ratio $E/\sigma_0$ by directly accessible geometrical factors, and so are the stress levels $\sigma_{x,i}$ close to the edges of the patterned area. However, as aforementioned, in the 2D inhomogeneous situation, one does not dispose of such direct relations. The value of the ratio $E/\sigma_{x,1}$ in the unpatterened region close to the middle of the edge is therefore extracted from the relation
\begin{equation}
\ell_{x,1}=\sqrt{\frac{E t_1^2}{12(1-\nu^2)\sigma_{x,1}}},
\label{eq:l1relation}
\end{equation}
where a value of $\nu=0.27$ is assumed~\cite{Etilde}. FEM simulations of the stress distribution $\sigma_{x}$ are then performed for various values of $\sigma_0$ in order to reproduce the value obtained from (\ref{eq:l1relation}). Let us note that, for highly stressed membranes, the level of stress released outside the patterned area is essentially independent of the value of the elasticity modulus, and, as in Eq.~(\ref{eq:sigmax}), only a function of $\sigma_0$ and the geometry. This implies that simulations performed with approximate values of $E$ and $\sigma_0$ do provide reliable estimates of the $E/\sigma_0$ ratio. We test this method for determining the $E/\sigma_0$ ratio in Sec.~\ref{sec:results} by applying it to different samples from each of the three batches introduced in Sec.~\ref{sec:samples}.

\subsection{Vibrational characterization}

\begin{figure}[h]
\includegraphics[width=0.45\columnwidth]{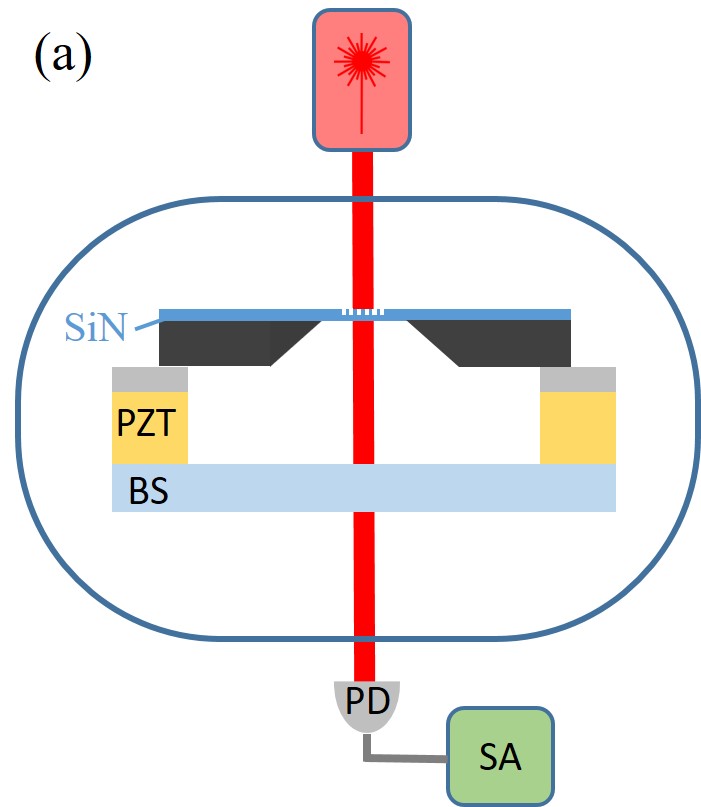}
\includegraphics[width=0.54\columnwidth]{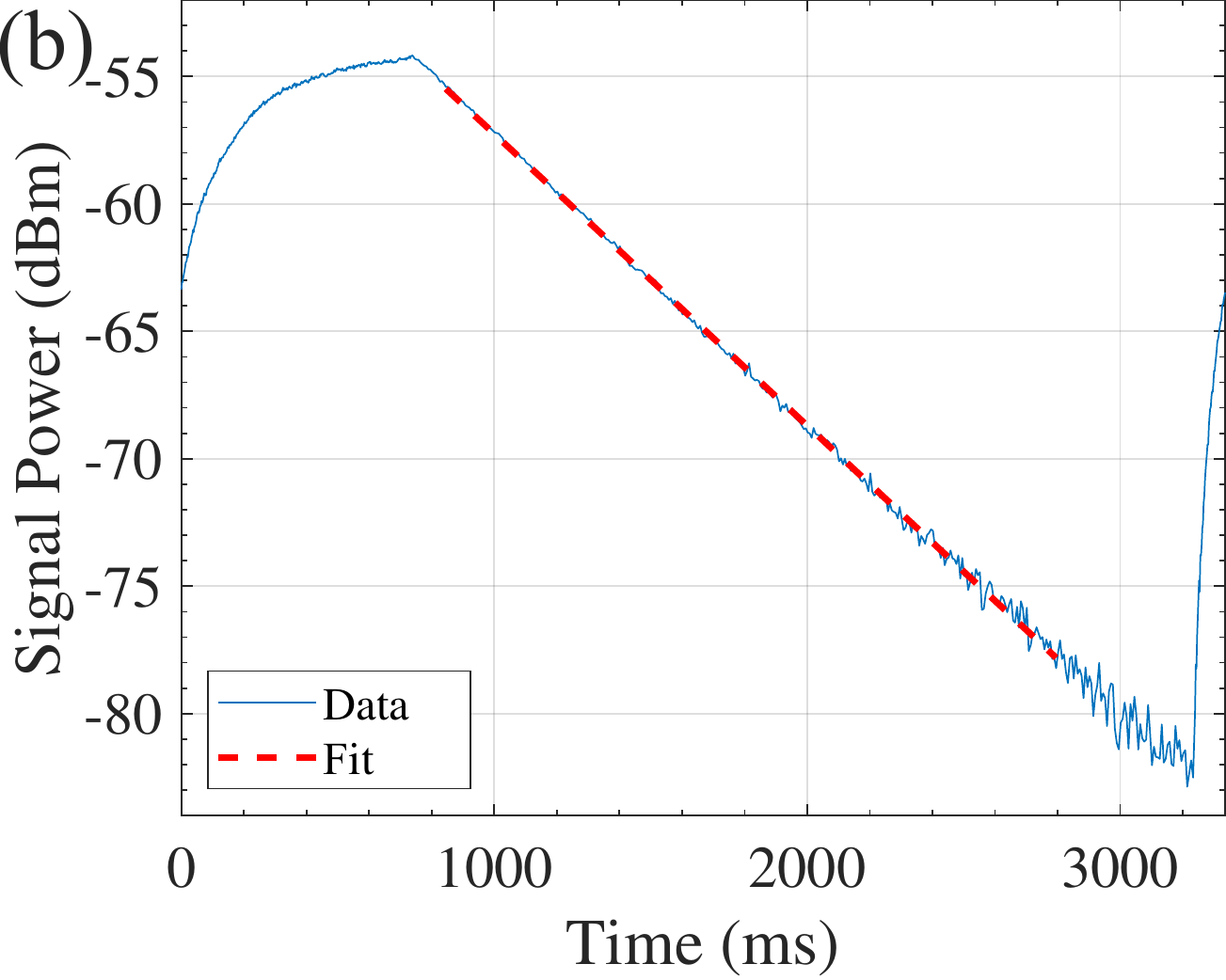}
\caption{(a) Schematic of the setup used for the vibrational characterization of the sample (PZT: piezoelectric transducer, BS: beamsplitter, PD: photodiode, SA: spectrum analyzer. (b) Example of a mechanical ringdown measurement of sample A1's (1,1) mode at 301.951 kHz, yielding a ringdown time of 759 ms, corresponding to a Q-factor of $7.2\times 10^5$.}
\label{fig:vibration}
\end{figure}

The frequencies and quality factors of the lowest frequency vibrational modes of the samples were determined in a low-pressure environment ($\sim 10^{-6}$ mbar) using standard optical interferometry and mechanical ringdown techniques, as described in~\cite{Nair2017,Naesby2017}. In brief, the sample is inserted in a vacuum chamber where the membrane, resting on the corners of its Si frame on a circular mount, forms together with a 50:50 beamsplitter a linear Fabry-Perot interferometer. This interferometer is illuminated by monochromatic light ($\sim 900$ nm) from an external cavity diode laser. The light transmitted by the interferometer is recorded on a photodiode and the signal sent to a low resolution bandwidth spectrum analyzer. A piezoelectric transducer between the sample and the beamsplitter allows for tuning the interferometer length and for resonantly exciting the vibrational modes. A schematic of the setup and an example of mechanical ringdown measurement for sample A1 are given in Fig.~\ref{fig:vibration}. After a prolonged resonant excitation at the mechanical resonance frequency the excitation is abruptly switched off. An exponential fit to the amplitude decay then gives the amplitude decay time and, thereby, the quality factor of the resonance whose value and uncertainty are obtained by averaging over 6 measurements (yielding $Q\sim (7.1\pm 0.1)\times 10^5$ for the example of Fig.~\ref{fig:vibration}).

\begin{figure}[h]
\includegraphics[width=\columnwidth]{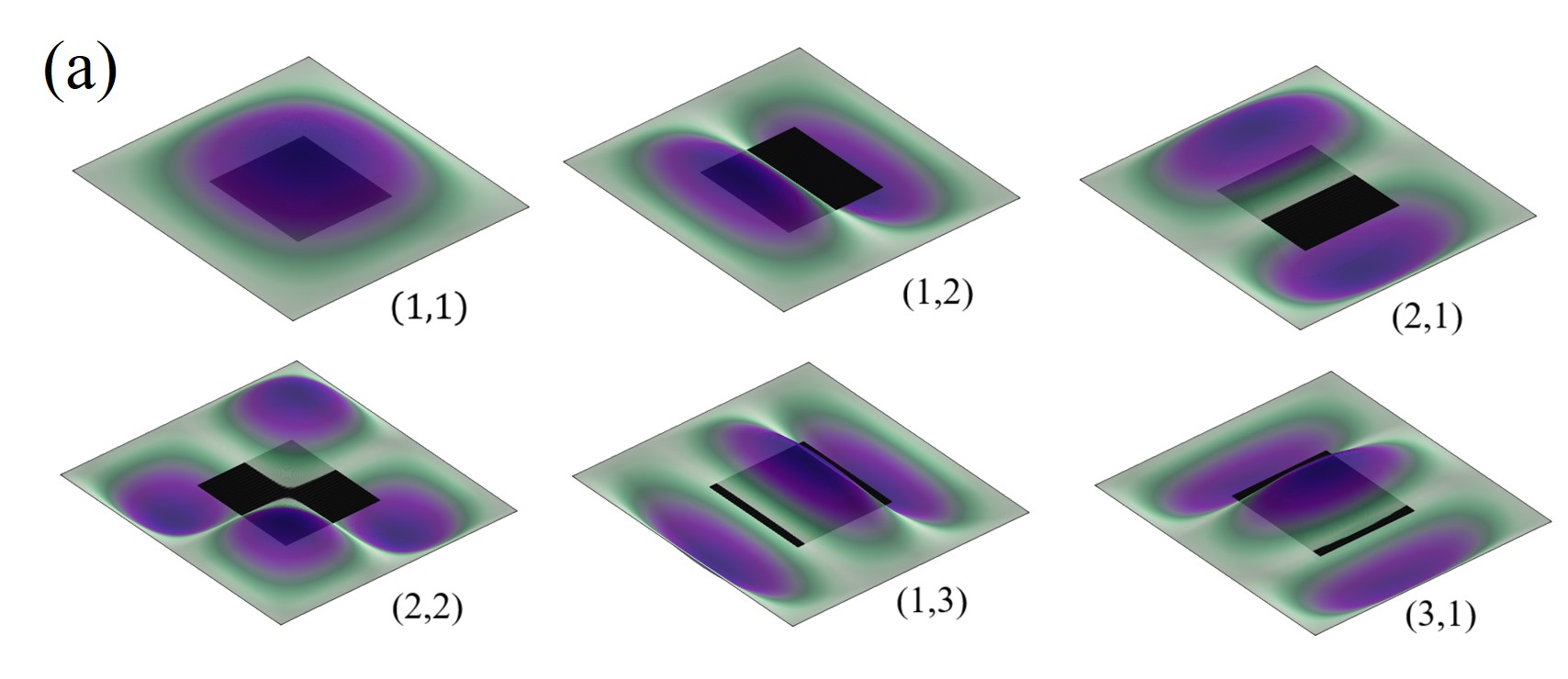}
\includegraphics[width=\columnwidth]{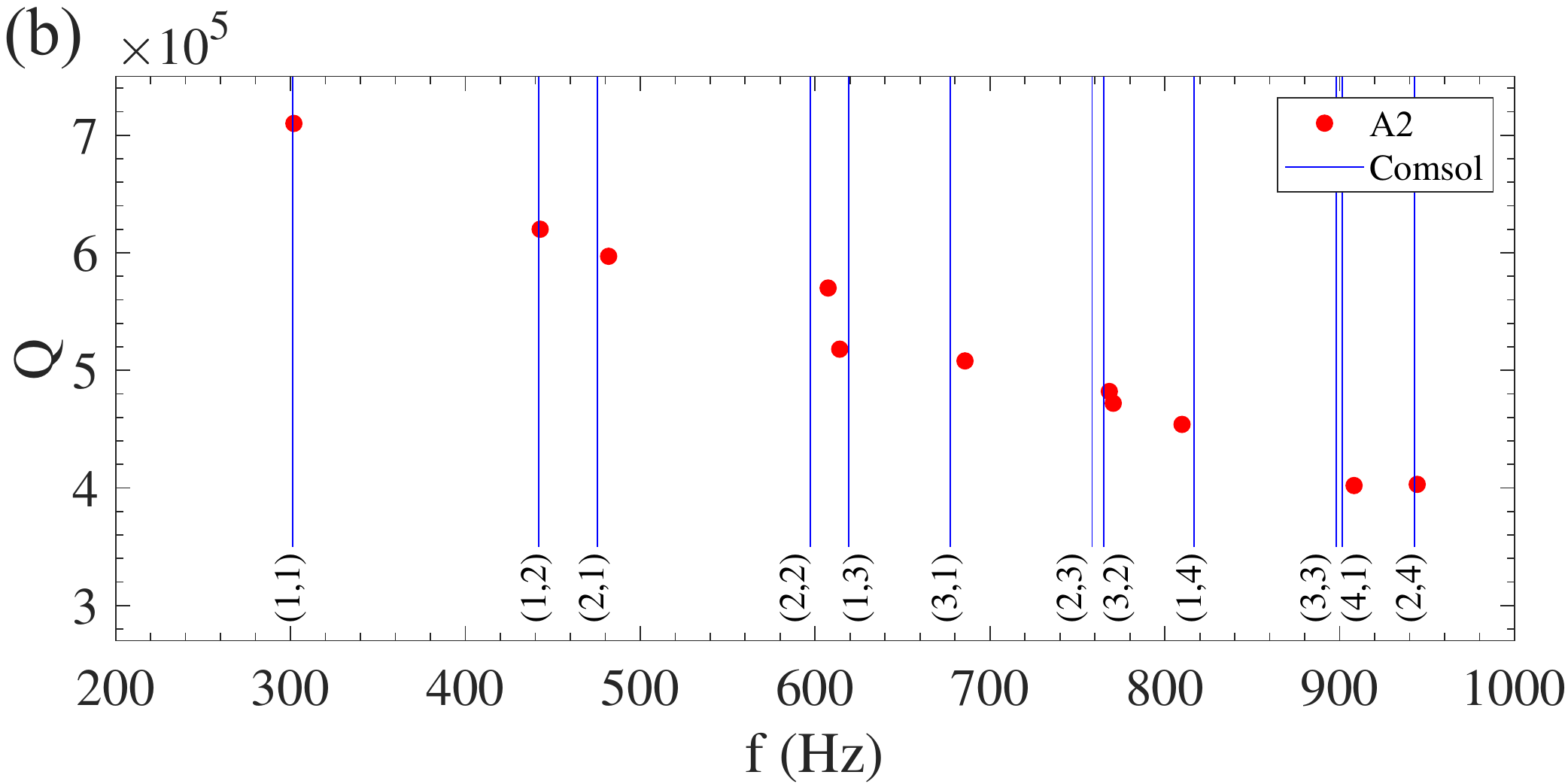}
\caption{(a) FEM-simulated vibrational mode shapes for the 6 lowest frequency modes. (b) Measured mechanical resonance frequencies and quality factors for sample A1. The vertical lines show the corresponding resonance frequencies obtained by FEM for $\rho/\sigma_0=21.6$~kg/m$^3$/MPa. The black squares indicate the grating-patterned area.}
\label{fig:frequencies}
\end{figure}

For each sample a vibrational eigenmode analysis is performed in COMSOL, based on the static structural analysis described in the previous section which uses the geometrical grating parameters experimentally determined from the AFM scans as input and using clamped boundary conditions. For the levels of stress considered in this work the tensile stress contribution largely dominates over the flexural one, such that the frequencies of the vibrational modes are essentially independent of the elasticity modulus $E$ and primarily determined by the ratio $\rho/\sigma_0$ of the film density $\rho$ and the level of prestress $\sigma_0$. For an unpatterned, square and homogeneous membrane the transverse eigenmodes are two-dimensional standing waves with eigenfrequencies given by \begin{equation}f_{m,n}=\sqrt{\frac{\sigma_0}{\rho}}\frac{1}{2a}\sqrt{m^2+n^2}\hspace{0.5cm}(m,n\in \mathbb{N}^*).\end{equation} For a membrane patterened with a square grating, the simulations show that the modefunctions are close to those of square drummodes (see Fig.~\ref{fig:frequencies}(a)). However, the eigenfrequencies are shifted from square drum eigenfrequencies, both due to the inhomogeneous decrease in mass after etching and the modification of the stress distribution. The inhomogeneous resulting stress distribution of the patterned films furthermore removes the degeneracy of modes with equal $m+n$. Nevertheless, the eigenfrequencies are still essentially determined by the ratio $\rho/\sigma_0$ (and, of course, the grating geometrical parameters). The ratio $E/\sigma_0$ is varied in the simulations so as to obtain the best match with the experimentally measured frequencies. Figure~\ref{fig:frequencies}(b) shows as an example experimentally measured frequencies and quality factors of the first lowest vibrational modes for sample A1, together with the frequencies predicted by FEM for a ratio $\rho/\sigma_0=21.6$~kg/m$^3$/MPa. We test this method for determining the $\rho/\sigma_0$ ratio in the next Section by applying it to different samples from each of the three batches introduced in Sec.~\ref{sec:samples}.

\subsection{Point deflection spectroscopy}

\begin{figure}[h]
\includegraphics[width=\columnwidth]{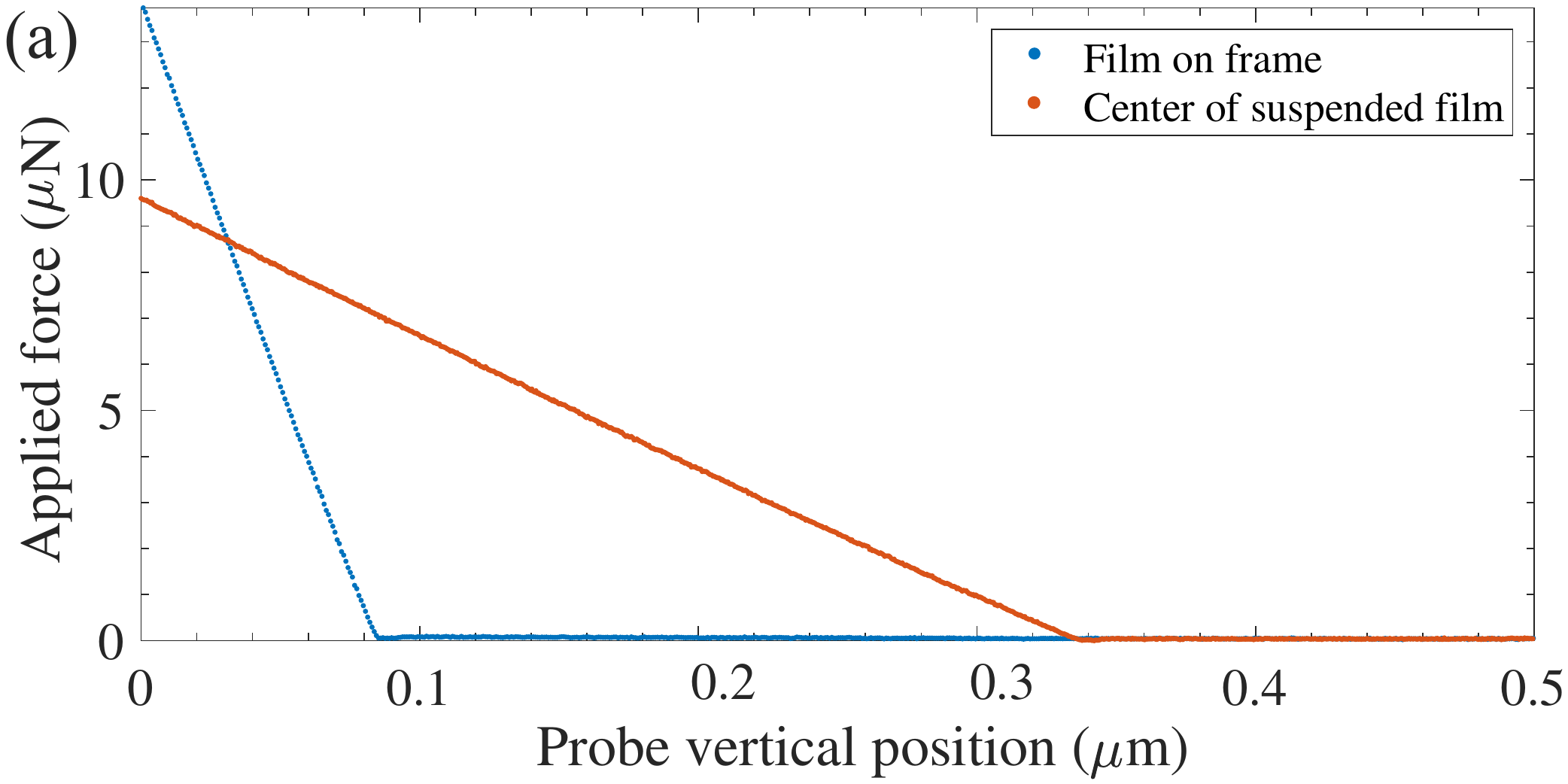}\\
\includegraphics[width=\columnwidth]{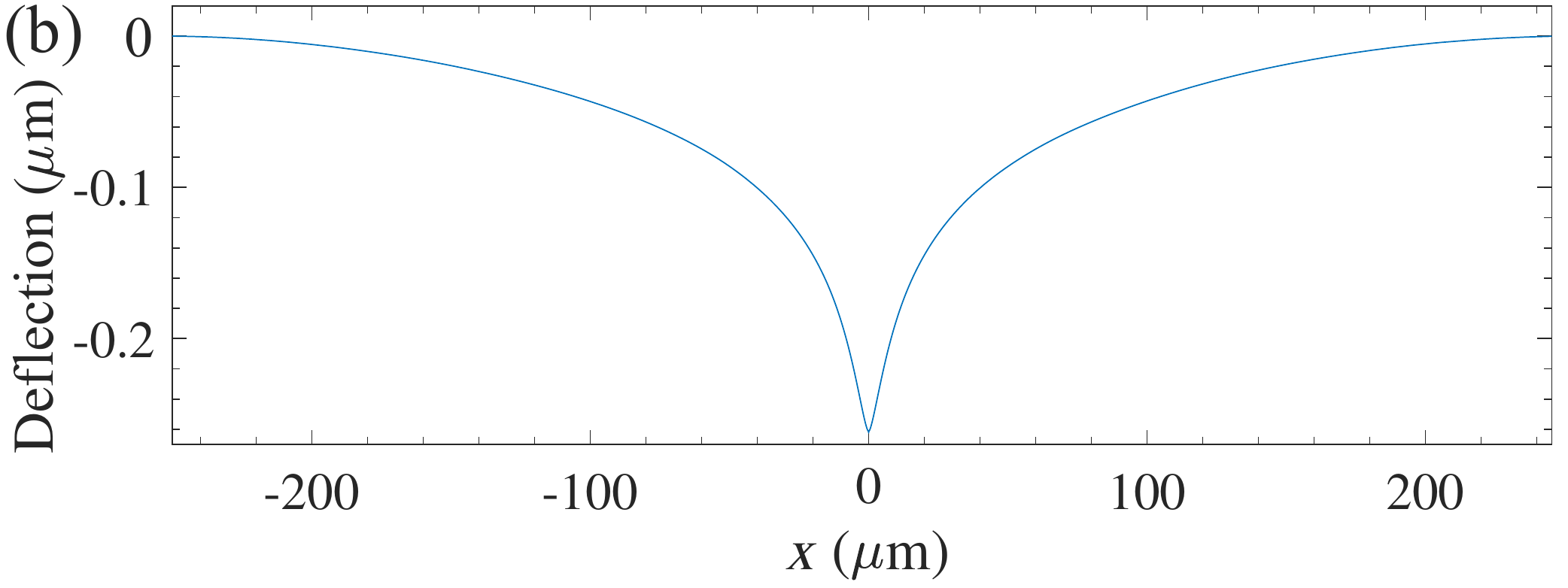}
\caption{(a) Measured deflection-force curve of the suspended (red) and non-suspended (blue) film of an unpatterned B membrane. (b) FEM-simulated deflection profile for a 10 $\mu$N force applied at the center of the membrane.}
\label{fig:contact}
\end{figure}

In order to measure the residual tensile stress level of unpatterned samples, standard AFM contact force spectroscopy is performed. Deflection-force curves, such as shown in Fig.~\ref{fig:contact}(a), are obtained by measuring the deflection caused by a small force ($\sim 10$~$\mu$N) applied with a Brucker RTESPA-525 probe, first on the film deposited on the Si frame and then at the center of the suspended film. The AFM tip spring constant $K_p = 175.4$~N/m is obtained from Brucker calibration probes specified with a $\pm 5\%$ uncertainty. Assuming the frame to be rigid, the spring constants, $K_p$ and $K$, measured in presence and absence of a rigid support, respectively, yield the suspended thin film spring constant $K_m=1/(K^{-1}-K_p^{-1})$. For each sample, a force-deflection curve is measured at the 4 edges and repeated 4 times at the membrane center.

In order to relate the extracted spring constant to the pretension level $\sigma_0$ FEM simulations are performed for various values of $\sigma_0$ and an approximate guess for $E$ given by the previously determined $E/\sigma_0$. The FEM simulations are performed by applying a localized force 10 $\mu$N on a 10 nm-radius spot at the center of the membrane and simulating its deflection. Again, the exact value of $E$ is not critical for the determination of $\sigma_0$ via the spring constant. The results of the simulated deflection profile for an unpatterned B batch membrane is shown in Fig.~\ref{fig:contact}(b). The extracted pretension level for membranes from the 3 batches are given in Tab.~\ref{tab:contact} in the next section.

For unpatterned membranes, the FEM simulation results were cross-checked against the semi-analytical predictions of~\cite{Jozwik2004,Martins2009}, which give the relationship between the maximum deflection $d$ of a square, pretensioned and homogeneous membrane and the concentrated load $F$ applied at its center. For small deflections, this relationship reads~\cite{Jozwik2004,Martins2009}
\begin{equation}
F=\frac{t\sigma_0}{\alpha\beta^2k_0^2g(k_0)}d+\frac{C_0}{1-\nu^2}\left(1+\frac{\eta}{g(k_0)}\right)\frac{tE}{a^2}d^3,
\end{equation}
where $\alpha=5.61\times 10^{-3}$, $\beta=1.86$, $C_0=6.08$, $\eta=1.9\times 10^{-2}$ are constants determined by FEM, $k_0$ is defined as
\begin{equation}
k_0^2=\frac{12(1-\nu^2)}{\beta^2}\left(\frac{a}{t}\right)^2\frac{\sigma_0}{E},
\end{equation}
and the function $g(k_0)$ is given by
\begin{align}
\nonumber g(k_0)&=\frac{8}{k_0^2}\left[\frac{K_1(k_0)-\frac{1}{k_0}}{I_1(k_0)}(I_0(k_0)+1)+K_0(k_0)+\ln\left(\frac{k_0}{2}\right)+\gamma\right]\\
&=\frac{8}{k_0^2}g'(k_0),
\end{align}
where $K_0$ and $K_1$ are Bessel functions of the second kind, $I_0$ and $I_1$ are modified Bessel functions of the first kind, and $\gamma$ is Euler's constant.


\section{Experimental results and analysis}\label{sec:results}

\subsection{Low-stress thick membranes}

\begin{figure}[h]
\centering
\includegraphics[width=\columnwidth]{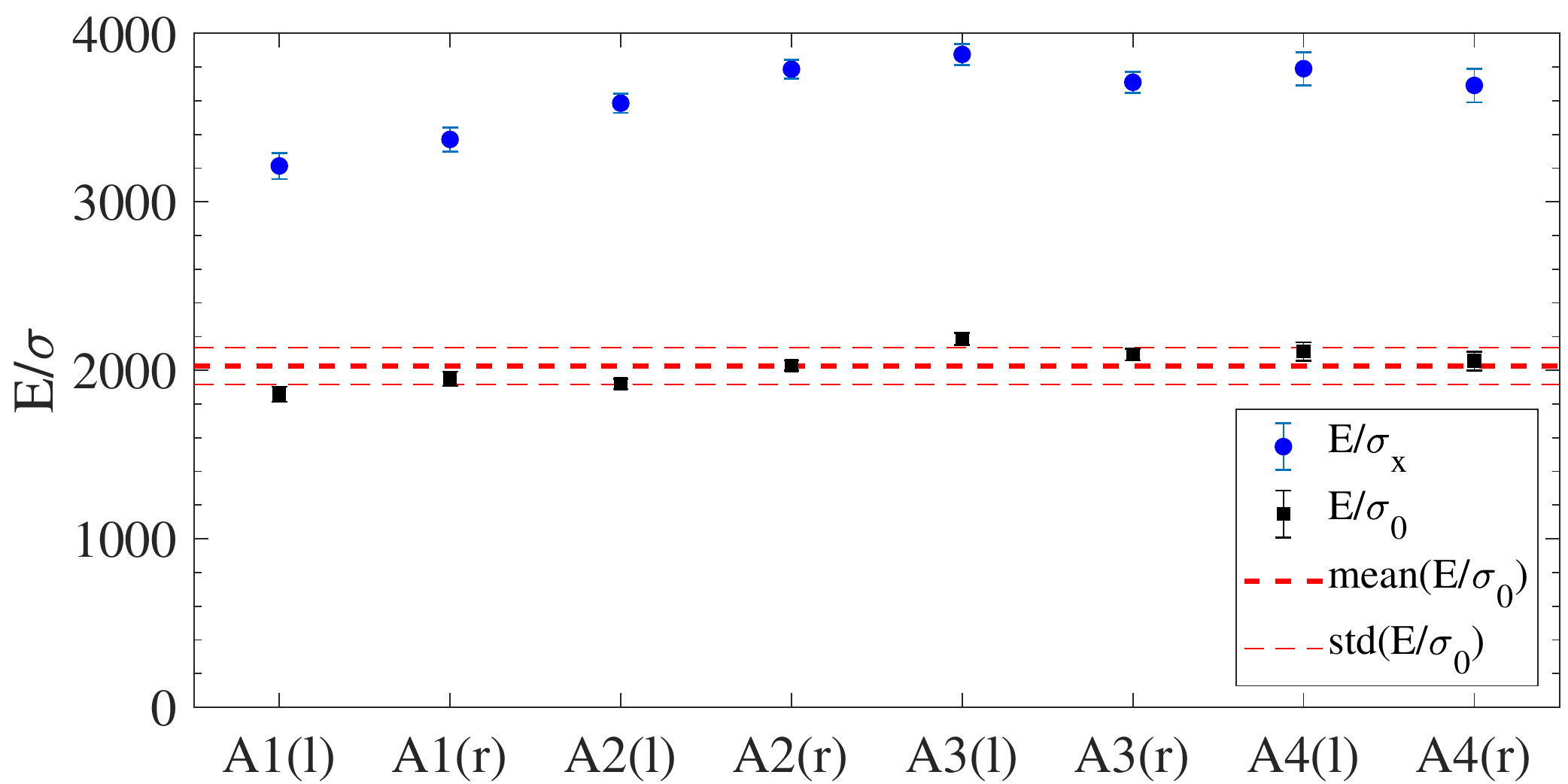}
\caption{Inferred $E/\sigma_x$ (blue circles) and $E/\sigma_0$ (black squares) ratios for batch A membranes. The thick and thin dashed red lines indicate the mean and the $\pm 1\sigma$ standard deviations, respectively.} 
\label{fig:ratiosJs}
\end{figure}

We first report on the characterization of the low stress, 315 nm-thick membranes from batch A, whose geometrical parameters are close to identical to each other. Table~\ref{tab:A} shows the measured deflection lengths at the center of the first finger edges resulting from AFM scans in the $x$-direction such as the one depicted in Fig.~\ref{fig:afm}. The deflections measured at the first and last fingers are arbitrarily denoted by \textit{left} (l) and \textit{right} (r). As explained in Sec.~\ref{sec:2Dmethod}, the value of $\ell_{x,1}$ for each scan is then used to infer the value of the ratio $E/\sigma_x$ in the stress-released region and FEM simulations are performed to infer what value of the ratio $E/\sigma_0$ is required to reproduce this value of $E/\sigma_x$. Both ratios are plotted in Fig.~\ref{fig:ratiosJs}. Averaging these results yields a quite accurate value of $2024\pm 109$ for the ratio $E/\sigma_0$.

\begin{table}[h]
\caption{\label{tab:A}Measured deflection lengths $\ell_{x,1}$ (in $\mu$m) for batch A membranes. }
\begin{ruledtabular}
\begin{tabular}{ccccc} 
Sample & A1 & A2 & A3 & A4\\
$\ell_{x,1}$ (l) & 5.31(6) & 5.62(4) & 5.84(4) & 5.78(8)\\
$\ell_{x,1}$ (r) & 5.45(6) & 5.75(4) & 5.74(4) & 5.68(8)
\end{tabular}
\end{ruledtabular}
\end{table}

\begin{figure}[h]
\centering
\includegraphics[width=\columnwidth]{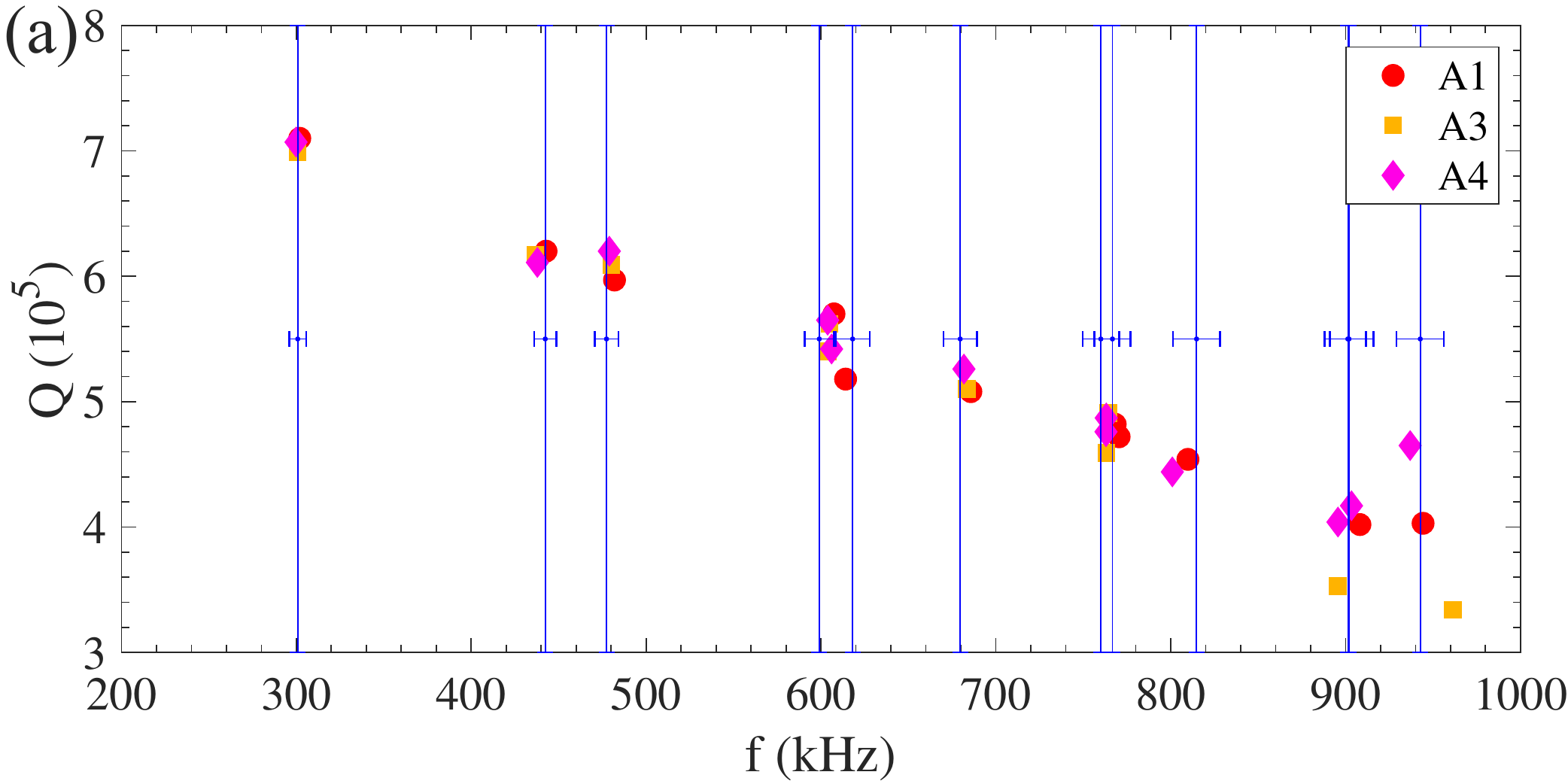}
\includegraphics[width=\columnwidth]{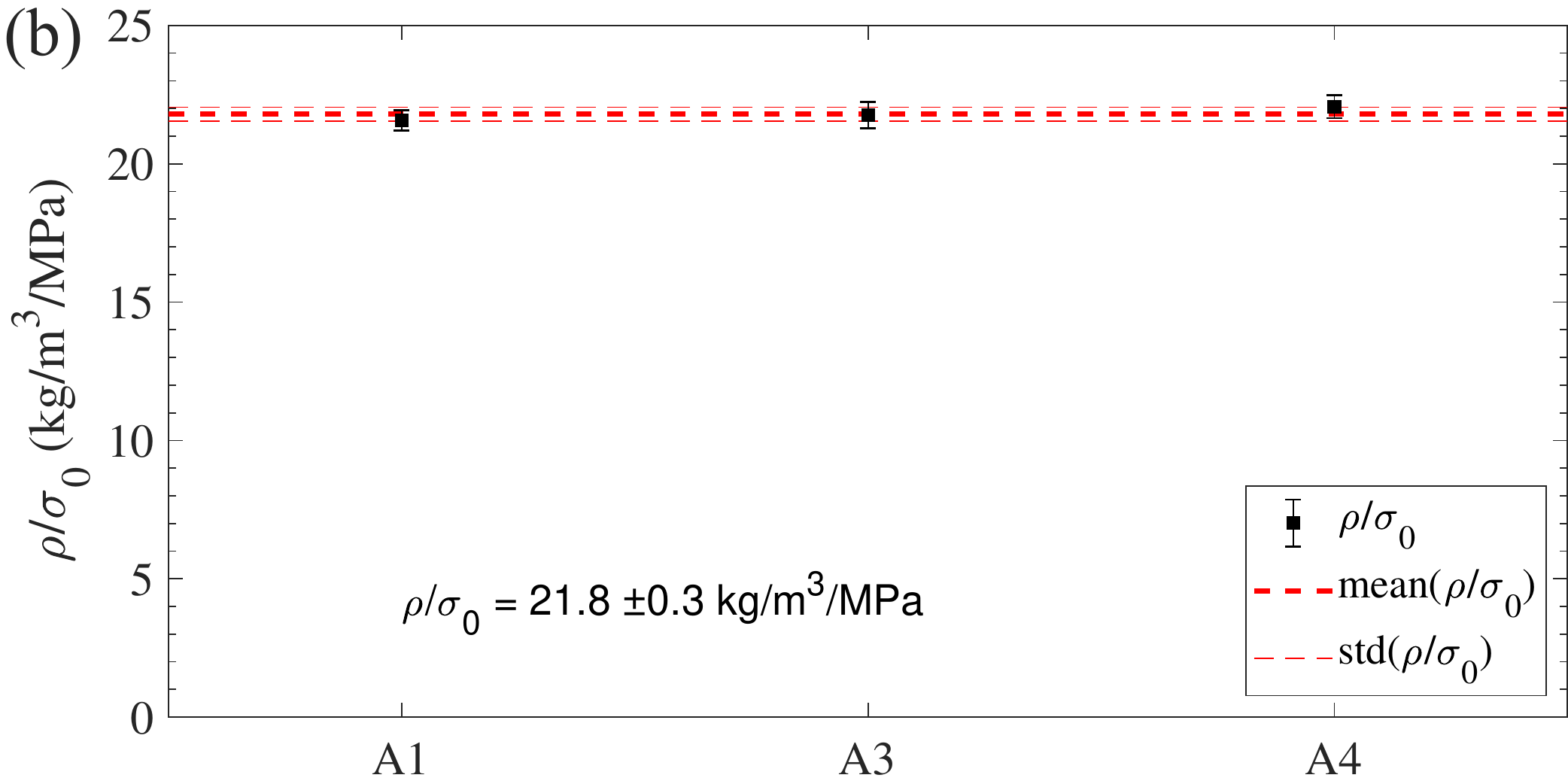}
\caption{(a) Measured mechanical resonance frequencies and quality factors for membranes A1, A3 and A4. The vertical lines show the predictions of the FEM simulations after optimizing the ratio $\rho/\sigma_0$. (b) Inferred ratios $\rho/\sigma_0$. The thick and thin dashed red lines indicate the mean and the $\pm 1\sigma$ standard deviation values, respectively.}
\label{fig:QvsfJs}
\end{figure}

Figure~\ref{fig:QvsfJs} shows the measured resonance frequencies and associated quality factors of the lowest vibrational modes of batch A membranes. A high degree of homogeneity is again observed for these membranes, as well as high Qs ($\sim 7\times 10^5$ for the fundamental modes at $\sim 300 $ kHz). These Q-values are in line with has been reported with (unpatterned) membranes from the same producent with similar sizes and thicknesses~\cite{Wilson2009,Wilsonthesis}. This indicates that the patterning of even a large fraction of the area of the membrane with a grating structure does not seem to appreciably degrade the mechanical quality of the membranes. One should note, though, that a slight decrease of the Q-factors with frequency is systematically observed, in contrast with what is typically observed with unpatterned membranes~\cite{Wilson2009}. Since we did not dispose of unpatterned membranes for this batch, we have not been able to test whether this trend is caused by the patterning or is due to the relatively high thickness of the bare films. Figure~\ref{fig:QvsfJs} also shows that the measured frequencies are in good agreement with the FEM predictions after optimizing the average ratio $\rho/\sigma_0=21.8\pm 0.3$ kg/m$^3$/MPa.

\subsection{Low-stress thin membranes}

\begin{figure}[h]
\centering
\includegraphics[width=\columnwidth]{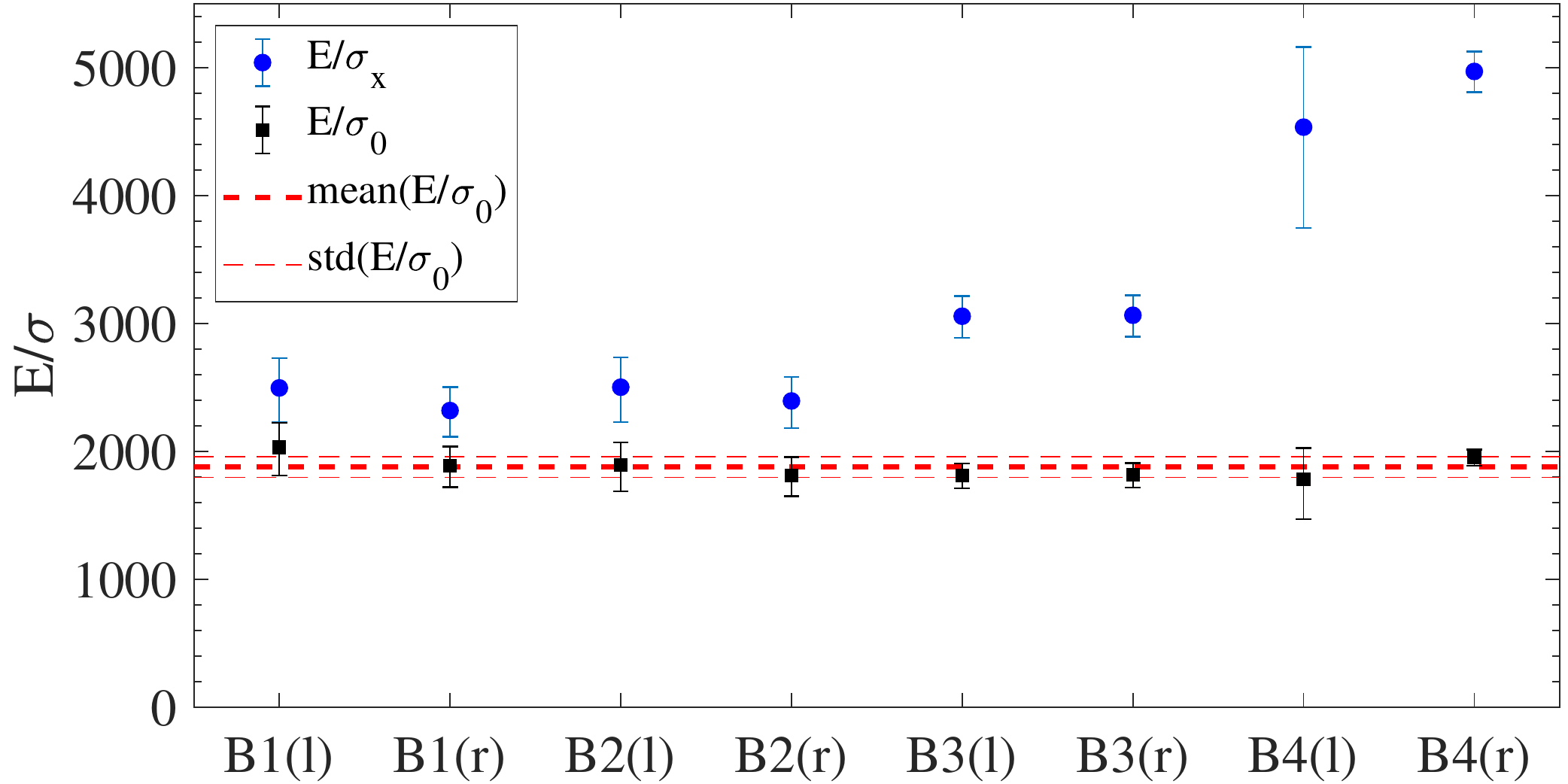}
\caption{Inferred $E/\sigma_x$ and $E/\sigma_0$ ratios for batch B membranes.}
\label{fig:ratiosHs}
\end{figure}

We now turn to the results of the characterization of batch B membranes, which possess a similar "low" level of pretension as batch A membranes, but are thinner (200~nm). In addition, the fabrication method is different, as the gratings are patterned on the already suspended SiN films using a combination of EBL and etching~\cite{Nair2019}. Last, the grating finger depth of the studied samples was varied by a factor of almost 2, thus substantially changing the level of stress released after patterning. Table~\ref{tab:B} reports the deflection lengths $\ell_{x,1}$ measured for 4 samples with different finger depths. A clear increase with $h$ of the deflection length in the unpatterned area is observed, concommitantly with the decrease of $\sigma_x$ in this region. The corresponding inferred values of $E/\sigma_x$ are plotted in Fig.~\ref{fig:ratiosHs}, together with the extracted values of the ratio $E/\sigma_0$. While the value of $E/\sigma_x$ now significantly varies from sample to sample due to the difference in the finger depth, the values of the ratio $E/\sigma_0$ are observed to be fairly constant, in agreement with the reasonable expectation that membranes from the same fabrication batch possess the same elasticity and level of pretension. Averaging over the 4 samples gives a value of $1879\pm 81$ for the $E/\sigma_0$ ratio of the batch B membranes.

\begin{table}[h]
\caption{\label{tab:B}Measured finger depths $h$ (in nm) and deflection lengths $\ell_{x,1}$ (in $\mu$m) for batch B membranes. The number in parenthesis indicate the $\pm 1\sigma$ standard deviation value.}
\begin{ruledtabular}
\begin{tabular}{ccccc} 
Sample & B1 & B2 & B3 & B4\\
$h$ & 98(5) & 106(4) &144(1) & 179(2)\\
$\ell_{x,1}$ (l) & 3.0(2) & 3.1(2) & 3.4(1) & 4.1(3)\\
$\ell_{x,1}$ (r) & 3.0(1) & 3.1(1) & 3.4(1) & 4.3(1)
\end{tabular}
\end{ruledtabular}
\end{table}

\begin{figure}[h]
\centering
\includegraphics[width=\columnwidth]{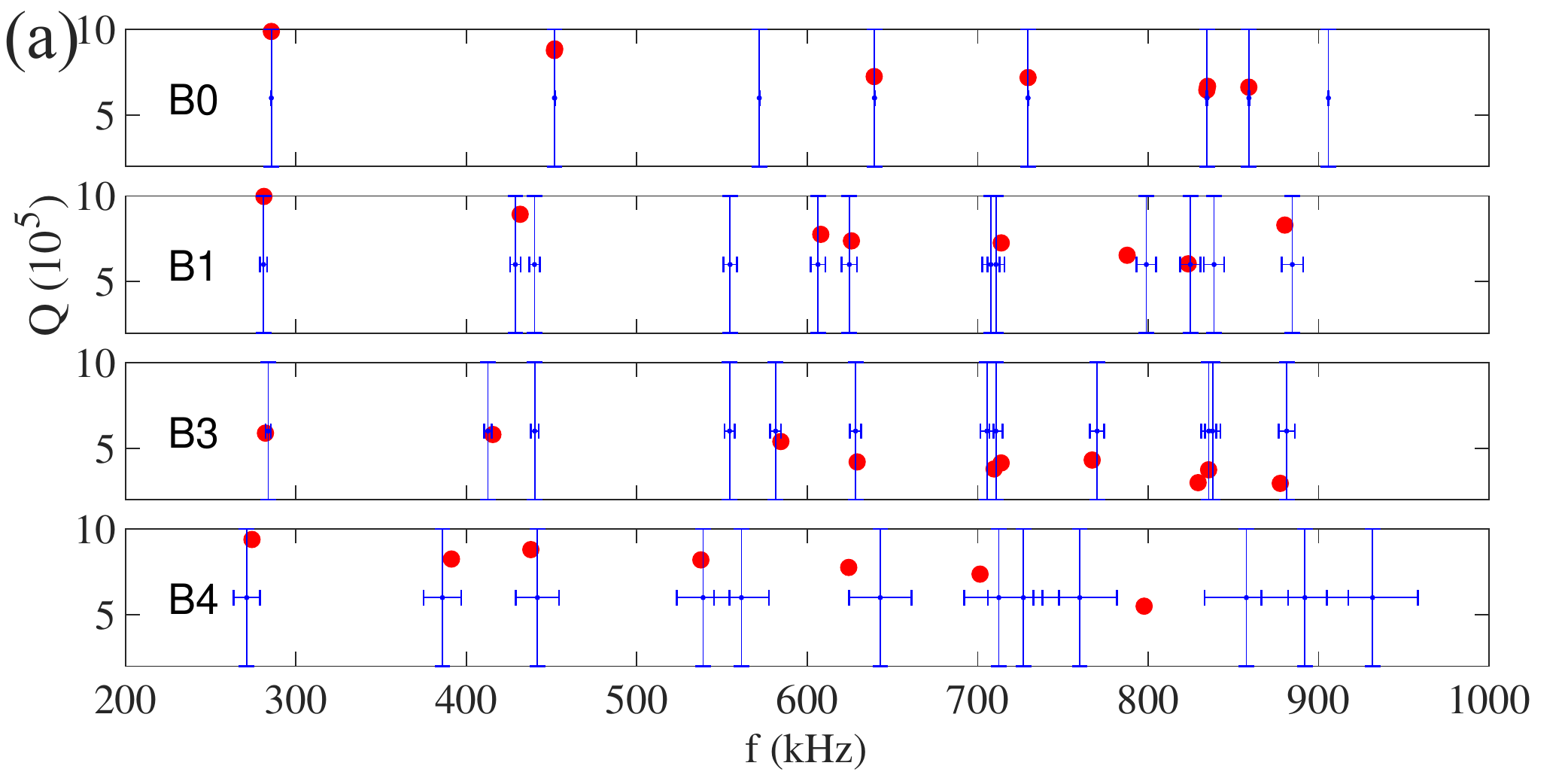}
\includegraphics[width=\columnwidth]{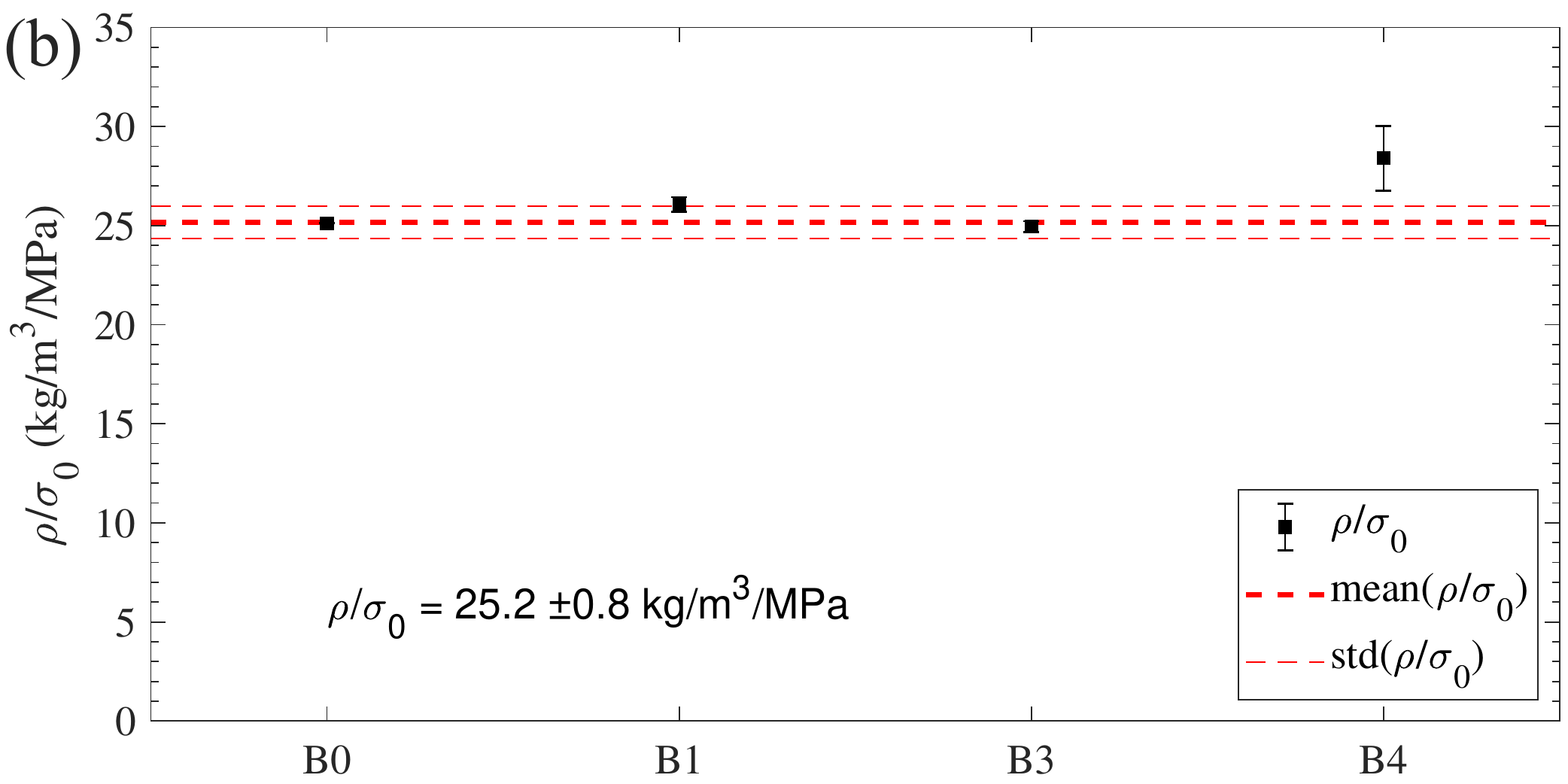}
\caption{(a) Measured mechanical resonance frequencies and quality factors for unpatterned membrane B0 and patterned membranes B1, B3, B4 (see Table~\ref{tab:B}).  The vertical lines show the predictions using FEM after optimizing the ratio $\rho/\sigma_0$. (b) Inferred ratios $\rho/\sigma_0$. The thick and thin dashed red lines indicate the mean and the $\pm 1\sigma$ standard deviation values, respectively.}
\label{fig:QvsfHs}
\end{figure}

Figure~\ref{fig:QvsfHs} shows the mechanical resonance frequencies and quality factors measured for three of these membranes, as well as for an unpatterned membrane from the same fabrication batch (B0). Again, high Q-factors in the $10^5-10^6$ range are observed for the patterned membranes, with a similar slight decrease with frequency as observed with the batch A membranes, and at the exception of B2, comparable with the Qs of the unpatterned membrane B0. Also shown in Fig.~\ref{fig:QvsfHs} are the predictions using FEM after optimization of the ratio $\rho/\sigma_0$. An increase in the frequency separation between modes with the same $m+n$ is clearly observed when increasing $h$ for the patterned membranes, as a result of the increased difference in the tensile stress in the $x$- and $y$-directions. The inferred values of the $\rho/\sigma_0$ ratio are also reported in Fig.~\ref{fig:QvsfHs}(b), yielding an average value of $25.2\pm 0.8$ kg/m$^3$/MPa for this batch. 

\subsection{High-stress thin membranes}
\label{sec:C}

Last, we report on the characterization of patterned membranes from batch C. These membranes have a similar thickness as the batch B membranes, as well as the same fabrication recipe. They are, however, stoichiometric Si$_3$N$_4$ films possessing a much higher pretension level ($\sim$GPa). In addition, the grating structures patterned on these membranes substantially vary in size, period, fill ratio and finger depth, thus incidentally providing a test of the robustness of our characterization method for determining the $E/\sigma_0$ and $\rho/\sigma_0$ ratios. The geometrical parameters of the patterned membranes investigated by AFM profilometry are reported in Table~\ref{tab:C}, together with the measured deflection lengths $\ell_{x,1}$. As previously, the measured deflection lengths are observed to be well correlated with the etching depths. The resulting $E/\sigma_{x,1}$ ratios follow the same trend, while the extracted $E/\sigma_0$ ratios again show quite consistent values, regardless of the grating parameters. A remarkably precise value of $253\pm 6$ is finally extracted for the $E/\sigma_0$ ratio by averaging over the measurements for these 4 samples.

\begin{table}[h]
\caption{\label{tab:C}Measured grating finger depths $h$ (in nm), period $\Lambda$ (in nm), fill ratio $f$, size $b$ (in $\mu$m) and deflection lengths $\ell_{x,1}$ (in $\mu$m) for the batch $C$ membranes whose structural characterization is reported in~Fig.~\ref{fig:ratiosCs}.}
\begin{ruledtabular}
\begin{tabular}{ccccccc} 
Sample & C1 & C2 & C3 & C4 & C5 & C6\\
$h$ & 86(1) & 90(2) & 90(1) & 153(1) & 93(3) & 92(1)\\
$\Lambda$ & 804(2) & 801(2) & 800(2) & 858(10) & 856(5) & 800(4)\\
$f$ & 0.47(2) & 0.54(1) & 0.57 (1) & 0.48(1) & 0.52(2) & 0.44(1)\\
$b$ & 200 & 200 & 100 & 200 & 200 & 400\\
$\ell_{x,1}$ (u) & 1.07(3) & 1.08(2) & 1.08(2) & 1.32(2) & - & -\\
$\ell_{x,1}$ (d) &  1.08(4) & 1.08(2) & - & - & - & -
\end{tabular}
\end{ruledtabular}
\end{table}

\begin{figure}[h]
\centering
\includegraphics[width=\columnwidth]{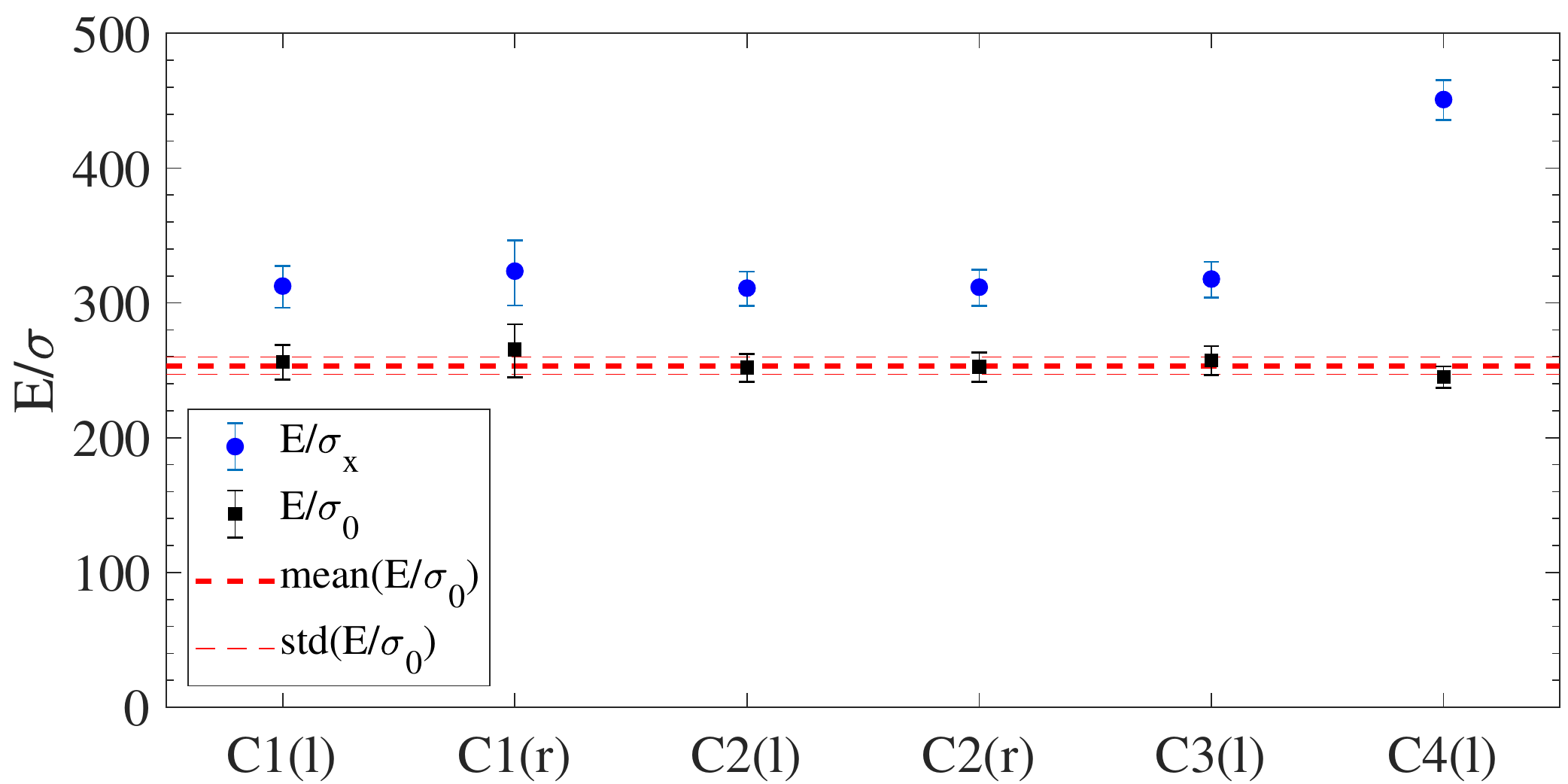}
\caption{Inferred $E/\sigma_x$ and $E/\sigma_0$ ratios for membranes C1-C4.}
\label{fig:ratiosCs}
\end{figure}

\begin{figure}[h]
\centering
\includegraphics[width=\columnwidth]{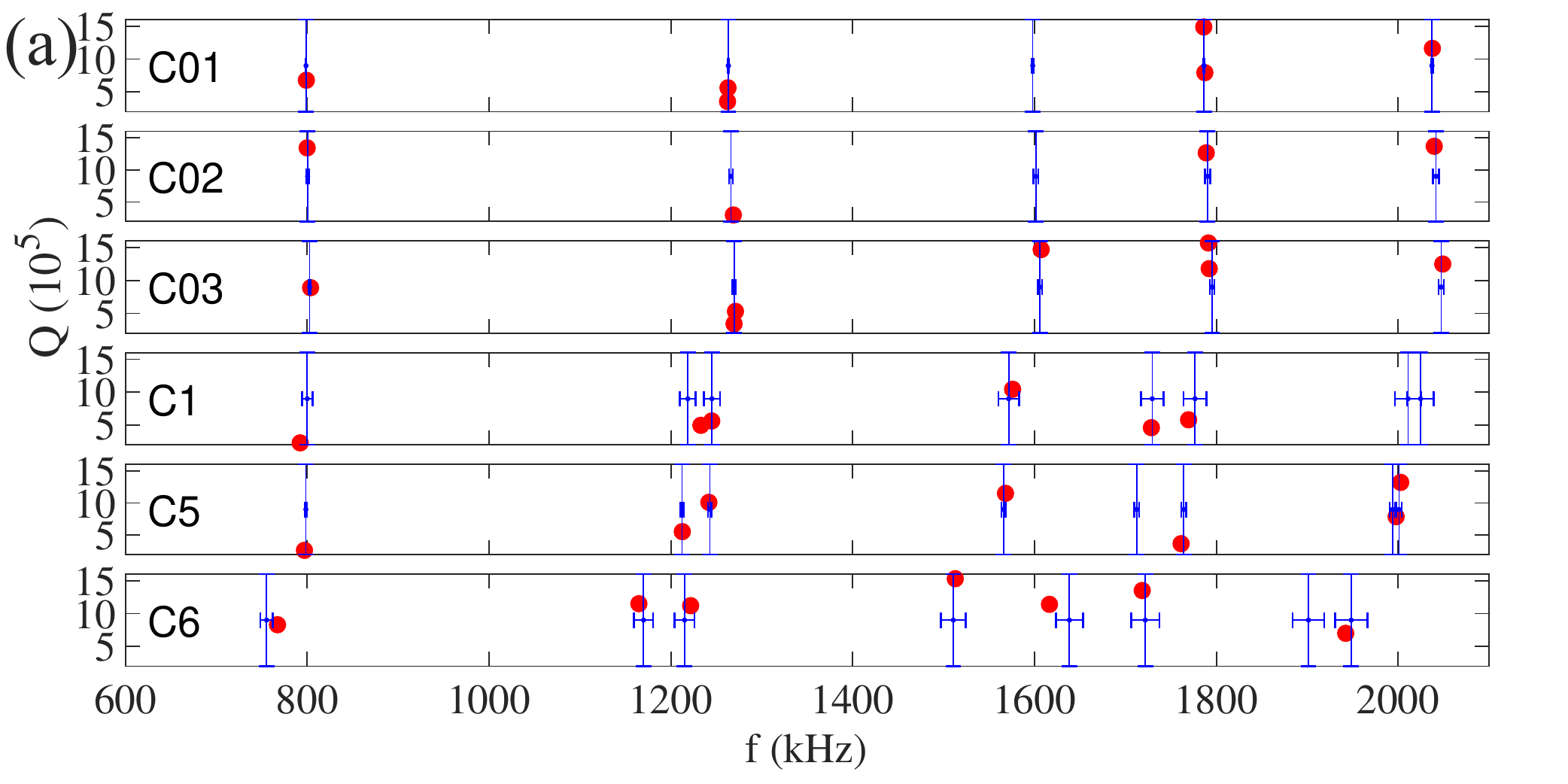}
\includegraphics[width=\columnwidth]{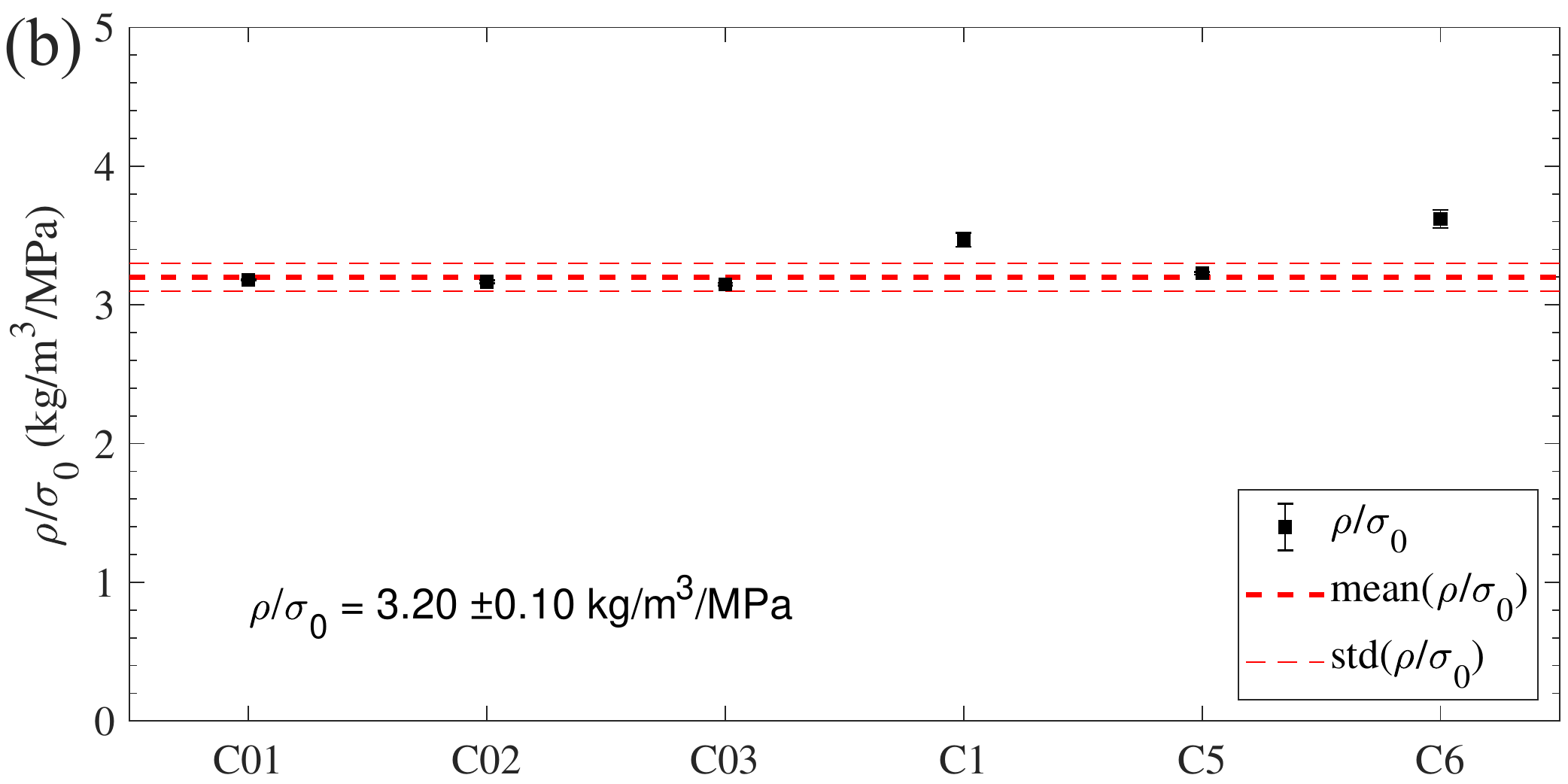}
\caption{(a) Measured mechanical resonance frequencies and quality factors for unpatterned membranes C01, C02, C03 and for patterned membranes C1, C5 and C6 (see Table~\ref{tab:C}). The vertical lines show the predictions using FEM after optimizing the ratio $\rho/\sigma_0$. (b) Inferred $\rho/\sigma_0$ ratios. The thick and thin dashed red lines indicate the mean and the $\pm 1\sigma$ standard deviation values, respectively.}
\label{fig:QvsfCs}
\end{figure}

Table~\ref{tab:C} shows the parameters of the high-stress membranes used in the vibrational measurements and whose characterization is reported in Fig.~\ref{fig:QvsfCs}. In addition to 3 membranes patterned with gratings possessing quite different sizes and parameters, the lowest vibrational frequencies and their associated Q-factors of an unpatterned membrane from the same fabrication batch are also shown. Again, good agreement is generally observed between measured and simulated frequencies when varying the ratio $\rho/\sigma_0$ in the simulations. Consistent values of the $\rho/\sigma_0$ ratio are finally obtained, yielding again a quite precise value of $3.20\pm0.10$ kg/m$^3$/MPa when averaging over the values of the 4 samples. 

It is also interesting to note that even the membrane with the largest grating whose dimension (400 $\mu$m) becomes comparable to the membrane size (500 $\mu$m) displays Qs similar to those of the unpatterned membranes, showing that fabricating large-area gratings on already suspended high-stress films using the recipe of \cite{Nair2019,Parthenopoulos2021} without degrading the mechanical quality of the resonators is possible.

\subsection{Deflection point spectroscopy and result summary}

The average spring constants determined by deflection spectroscopy for two unpatterned membranes of batches B and C and for two (patterned) membranes of batch A are reported in Table~\ref{tab:contact}. The values of the pretension stress level $\sigma_0$ extracted from FEM simulations in order to match these measured spring constants are also reported in the table and show similar pretension level for batches A and B, much lower than that of the stoichiometric films of batch C. The relative uncertainty in the spring constant/pretension level determination is mainly dominated by the 5\% uncertainty of the AFM calibration probes provided by the producent. These obtained values are also in agreement with wafer curvature measurements from Norcada yielding $156\pm 50$ and $1005\pm 50$ MPa for batches A and C, respectively. From the value of $\sigma_0$ and the average $E/\sigma_0$ and $\rho/\sigma_0$ ratios previously determined the values of the elasticity modulus and the density are then deduced and reported in Table~\ref{tab:contact}.

\begin{table}[h]
\caption{\label{tab:contact} Average spring constants $K_m$ measured by point deflection spectroscopy and extracted pretension levels $\sigma_0$ for the three batches. Average values of the elasticity modulus $E$ and the density $\rho$ deduced from $\sigma_0$ and the average $E/\sigma_0$ and $\rho/\sigma_0$ ratios previously determined. The numbers in parentheses indicate the corresponding relative uncertainties.}
\begin{ruledtabular}
\begin{tabular}{cccc}
Batch & A & B & C\\\hline
$K_m$ (N/m) & 42.9 ($\pm 10$\%) & 43.9 ($\pm 9$\%) & 244 ($\pm 5$\%) \\
$E/\sigma_0$ & 2024 ($\pm 5$\%) & 1879 ($\pm4$\%) & 253 ($\pm 2$\%)\\
$\rho/\sigma_0$ (kg/m$^3$/MPa) & 21.8 ($\pm 1$\%) & 25.2 ($\pm 3$\%) & 3.20 ($\pm3$\%) \\\hline
$\sigma_0$ (MPa) & 155 ($\pm 10$\%) & 135 ($\pm 9$\%) & 1030 ($\pm 7$\%)\\
$E$ (GPa) & 314 ($\pm11$\%) & 254 ($\pm10$\%) & 261 ($\pm 6$\%)\\
$\rho$ (kg/m$^3$) & 3379 ($\pm 10$\%) & 3402 ($\pm 9$\%) & 3296 ($\pm6$\%) 
\end{tabular}
\end{ruledtabular}
\end{table}


\section{Conclusion}\label{sec:conclusion}

A detailed investigation of the structural and vibrational properties of various photonic crytal SiN membranes was carried out. The lowest frequency resonances of the out-of-plane drummodes were experimentally determined by optical interferometry and shown to possess high-Q factors, which is promising for applications of these films in optomechanics and sensing. The comparison of the mechanical resonance frequencies with the results using FEM allows for accurately determining the ratio of the film density to its level of pretension before nanostructuring. Furthermore, by relating the nanostructuring-induced deformation of the film in the vicinity of the photonic crystal area measured by AFM scans, we show that the ratio of the elasticity modulus to the pretension level can be inferred in an accurate fashion. Last, using point deflection spectroscopy the pretension level of the films can be determined, finally yielding accurate estimates for the films' elasticity, density and residual stress levels. The methods applied here thus outline a way for noninvasively determining key material parameters of fragile nanostructured thin films, the knowledge of which is essential for their design, fabrication or performance optimization.


\begin{acknowledgments}
We acknowledge financial support from Independent Research Fund Denmark, as well as Bjarke R. Jeppesen for his assistance with AFM spectroscopy. We are grateful to Norcada for providing us with pretension level estimates for batches A and C. 
\end{acknowledgments}

\section*{Data Availability}
The data that support the findings of this study are available from the corresponding author upon request.





\begin{thebibliography}{99}
\bibitem{Etilde} If the Poisson ratio $\nu$ is not known, the method discussed only provides knowledge on the ratio $\tilde{E}=E/(1-\nu^2)$.
\end{thebibliography}
\end{document}